\documentclass{aa}
\usepackage{newtxtext,newtxmath}
\usepackage[T1]{fontenc}
\usepackage{ae,aecompl}
\usepackage{xcolor}
\bibpunct{(}{)}{;}{a}{}{,}
\usepackage[colorlinks=true, citecolor=blue]{hyperref}
\usepackage{multirow}
\makeatletter
\renewcommand*\aa@pageof{, page \thepage{} of \pageref*{LastPage}}
\makeatother
\usepackage{etoolbox}
\usepackage{array,booktabs}
\usepackage{graphicx}	
\usepackage{amsmath}	
\usepackage{verbatim}
\usepackage{units}
\usepackage{pdflscape}
\usepackage{svg}
\usepackage{bm}
\usepackage{stfloats}
\usepackage{float}
\usepackage{multicol}
\usepackage{hyperref}
\usepackage{multirow}
\usepackage{wasysym}
\usepackage{subcaption}

\newcommand{\degree}{\ensuremath{^\circ}}

\begin{document}
\title{Phenomenology and periodicity of radio emission from the stellar system AU\,Microscopii}

\author{S.~Bloot$^{1,2}$\thanks{Corresponding author \email{bloot@astron.nl}},
J.~R.~Callingham$^{1,3}$, H.~K.~Vedantham$^{1,2}$, R.~D.~Kavanagh$^{1}$, B.~J.~S.~Pope$^{4,5}$, J.~B.~Climent$^{6,7}$, J.~C.~Guirado$^{6,8}$, L.~Pe\~{n}a-Mo\~{n}ino$^{9}$, M.~P\'{e}rez-Torres$^{9,10,11}$}
\authorrunning{S.~Bloot et al.}
\institute{$^{1}$ASTRON, Netherlands Institute for Radio Astronomy, Oude Hoogeveensedijk 4, Dwingeloo, 7991\,PD, The Netherlands\\
$^{2}$Kapteyn Astronomical Institute, University of Groningen, P.O. Box 800, 9700 AV, Groningen, The Netherlands\\
$^{3}$Leiden Observatory, Leiden University, PO\,Box 9513, 2300 RA, Leiden, The Netherlands\\
$^{4}$School of Mathematics and Physics, The University of Queensland, St Lucia, QLD 4072, Australia\\
$^{5}$Centre for Astrophysics, University of Southern Queensland, West Street, Toowoomba, QLD 4350, Australia\\
$^{6}$Departament d'Astronomia i Astrof\'{i}sica, Universitat de Val\`{e}ncia, Burjassot, E-46100, Spain\\
$^{7}$Universidad Internacional de Valencia (VIU), Valencia, E-46002, Spain\\
$^{8}$Observatori Astronòmic, Universitat de València, Parc Científic, Paterna, E-46980, Spain\\
$^{9}$Instituto de Astrof\'isica de Andaluc\'ia (IAA-CSIC), Glorieta de la Astronom\'ia s/n, E-18008, Granada, Spain\\
$^{10}$Facultad de Ciencias, Universidad de Zaragoza, Pedro Cerbuna 12, E-50009, Zaragoza, Spain\\
$^{11}$School of Sciences, European University Cyprus, Diogenes street, Engomi, 
1516 Nicosia, Cyprus
}

\date{Received XXX; accepted 2023-12-13}

\label{firstpage}

\abstract{Stellar radio emission can measure a star's magnetic field strength and structure, plasma density, and dynamics, and the stellar wind pressure impinging on exoplanet atmospheres. However, properly interpreting the radio data often requires temporal baselines that cover the rotation of the stars, orbits of their planets, and any longer-term stellar activity cycles. Here we present our monitoring campaign on the young, active M\,dwarf \object{AU\,Microscopii} with the Australia Telescope Compact Array between 1.1 and 3.1\,GHz. With over 250 hours of observations, these data represent the longest radio monitoring campaign on a single main-sequence star to date. We find that AU\,Mic produces a wide variety of radio emission, for which we introduce a phenomenological classification scheme predicated on the polarisation fraction and time-frequency structure of the emission. Such a classification scheme is applicable to radio emission from other radio-bright stars. The six types of radio emission detected on AU\,Mic can be broadly categorised into five distinct types of bursts, and broadband quiescent emission. We find that the radio bursts are highly circularly polarised and periodic with the rotation period of the star, implying that the emission is beamed. It is therefore most likely produced by the electron cyclotron maser instability. We present a model to show that the observed pattern of emission can be explained by emission from auroral rings on the magnetic poles. 
The total intensity of the broadband emission is stochastic, but we show that its circular polarisation fraction is also periodic with the rotation of the star. Such a periodicity in the polarised fraction of emission has not been observed on an M\,dwarf before.
We present a qualitative model to describe the periodicity in the polarisation fraction of the broadband emission, using low-harmonic {gyromagnetic} emission. Using a simple qualitative model, we infer a magnetic obliquity of at least 20\degree ~from the observed variation in polarisation fraction. Finally, we show that the radio emission might be evolving on long timescales, hinting at a potential stellar magnetic activity cycle.}

\maketitle
\section{Introduction}
\label{sec:intro}
\object{AU\,Microscopii} (\object{AU\,Mic} hereafter) is an M1Ve-type star at a distance of 9.7\,pc that has been extensively studied across the electromagnetic spectrum due to its extreme activity, proximity, and unique planetary system \citep{1970PASP...82.1341K,2006A&A...460..695T,gaia, aumic_b,aumic_c}. The stellar system is thought to have formed relatively recently, with an estimated age of around 22\,Myr \citep{2014MNRAS.445.2169M}. The AU\,Mic system consists of the M\,dwarf primary with a debris disk, two confirmed planets \citep{aumic_b,aumic_c}, and two more candidate planets \citep{ttvs_aumic, zdi_aumic}. The two confirmed planets have orbital periods of 8.46\,days and 18.9\,days, which makes them ideal candidates for studying young planet atmospheres subjected to stellar activity.

AU\,Mic displays significant stellar activity, consistent with being a young low-mass system, with a high observed flare rate of two flares per day \citep{gilbert2022}. These flares, as well as the other effects of stellar activity, are important to characterise as they can strongly impact the evolution of the planets and their atmospheres \citep[see e.g.][for a review]{owen2019}, as well as inform us about the high-energy plasma components of the stellar corona \citep[e.g.][]{Gudel2002}. 
Stellar activity is most often studied from X-ray to optical wavelengths, but a new generation of sensitive radio telescopes has facilitated a resurgence in stellar radio observations \citep[e.g.][]{2020A&A...641A..90C, Zic_2020,callingham2021}. Radio observations of stars can provide direct measurements of the properties of the plasma around the star, as well as the magnetic field strength of the emitting body -- information that is difficult, or impossible, to obtain at other wavelengths \citep[e.g.][]{dulk1985,2006ApJ...637.1016O,2008ApJ...674.1078O}. For example, detections of few-minute-long radio bursts can be a direct probe of the propagation of coronal mass ejections, whereas longer-duration events could trace the trajectory of a plasmoid \citep[e.g][]{1950AuSRA...3..387W, 1985srph.book.....M,dulk1985}. Other radio bursts can measure the strength of the stellar magnetic field, from the large-scale dipole to local small-scale variations \citep[e.g.][]{2018ApJ...854....7L}. 

In the Solar System, we have two archetypal examples of bright emission at radio frequencies. The first example is the Sun, which produces a rich phenomenology of radio emission, including broadband quiescent emission and transient narrowband radio bursts \citep[e.g.][]{dulk1985}. The emission on the Sun is ultimately powered by magnetic reconnection in flares, and does not vary periodically with the rotation of the Sun, but it does evolve with the activity cycle of the Sun.
Emission that is a direct analogue to the solar radio emission has rarely been detected on other stars. A potential example was found by \citet{Zic_2020}, who detected a radio burst on Proxima Centauri that resembled a solar Type\,IV burst.

The other archetypal radio emitter in the Solar System is Jupiter. In contrast to the Sun, Jupiter produces periodic radio bursts, powered by magnetospheric processes that also cause optical and UV aurorae. We refer to these processes as auroral in the rest of this work. The cause of the coherent radio emission from Jupiter is twofold: (a) a breakdown of corotation between the plasma and magnetic field in the magnetosphere, causing radio emission that is periodic with the rotation of Jupiter, or (b) an interaction between the Jovian magnetic field and its moons (notably Io), producing radio emission that is periodic with the satellite's orbit \citep[e.g.][]{1998JGR...10320159Z}.

Since nature produces a continuum of objects between Solar and planetary mass scales, we expect to observe both Solar-like and Jovian-like emission from extrasolar systems. 
Jupiter-like auroral emission has been detected on brown dwarfs \citep[e.g.][]{hallinan2015,2016ApJ...818...24K,2022ApJ...932...21K}. This is not unexpected, as we expect most brown dwarfs to have magnetospheres with low plasma densities and high magnetic field strengths, similar to Jupiter. Although auroral radio emission is not seen on the Sun, it is possible that other stars, in particular stars with strong magnetic fields, may produce auroral emission as well. Early-type magnetic stars are indeed known to produce rotationally modulated pulsed emission similar to Jupiter \citep[e.g.][]{2017MNRAS.467.2820L,2019MNRAS.482L...4L,2020MNRAS.493.4657L}. In addition, there have been recent suggestions of auroral radio emission from some M\,dwarfs \citep[e.g.][]{villadsen2019,zic2019,callingham2021,crdra,Bastian_2022}. To confirm that the emission of an object is auroral in nature, a clear detection of periodicity, preferably with a known period in the system, is required.

The second type of radio emission detected from the Jupiter system is the Jupiter-Io interaction. An analogue of this process is predicted to be produced from stellar systems, with a star taking the place of Jupiter and a planet the place of Io \citep[e.g.][]{2007P&SS...55..598Z,2011A&A...531A..29H, 2013A&A...552A.119S,2016pmf..rept.....L,j1019,kavanagh2021,kavanagh2023}. At the time of writing, there have been several suggested detections of a star-planet interaction (SPI) in M\,dwarf systems \citep{j1019,perez-torres2021, crdra, pineda2023, yzceti2}, although none have been explicitly confirmed yet. A confirmation again requires a clear detection of periodicity, in this case at the orbital or synodic period.

Considering the strong magnetic fields required for auroral processes, either induced by a planet or powered by the star itself, to be detectable at gigahertz frequencies, M\,dwarfs are great targets to find evidence of these types of emission. Furthermore, M\,dwarfs often produce more detectable bright optical flares than the Sun \citep[e.g.][]{2020AJ....159...60G}. For these reasons, many M\,dwarfs have been studied at radio frequencies. \citet{1989AJ.....98..279F} and \citet{2005ApJ...626..486B} detected several M\,dwarfs at 4.8\,GHz, but were limited in their interpretation by the narrow bandwidth of their observations. More recently, \citet{villadsen2019} detected a total of 22 bursts on five different M\,dwarfs. \citet{crdra} detected three bursts on CR\,Dra in the longest monitoring campaign of an M\,dwarf with LOFAR. At those same frequencies, \citet{callingham2021} detected 19 M\,dwarfs in Stokes\,V. \citet{zic2019} detected rotationally modulated pulses on UV Ceti with ASKAP, with a strong periodicity at the rotation rate of the star. These results show that M\,dwarfs can produce highly circularly polarised radio bursts that last for more than an hour that show similarity to Jupiter's radio emission. 
Despite the number of radio detections of M\,dwarfs, no single M\,dwarf has ever been studied for enough time to fully explore the complete phenomenology of radio emission. To explicitly confirm the suggestion that M\,dwarfs are similar to Jupiter in terms of their radio emission, we require a clear detection of periodicity. Such a result can only be achieved with a large campaign on a single M\,dwarf, consisting of enough data to cover several rotations of the star.

Considering the previous detections of M\,dwarfs, AU\,Mic is an ideal target to observe at radio frequencies, based on both its magnetic field strength and its high activity level. The maximum magnetic field strength has been measured to be of the order of 1-2\,kG \citep{2020ApJ...902...43K, 2022MNRAS.512.5067K, zdi_aumic}, which would result in electron cyclotron maser emission (ECMI; see Section\,\ref{sec:structure}) at a frequency of 1-5\,GHz. The plasma density, estimated to be greater than $10^{11}$\,cm$^{-3}$ \citep{2002A&A...390..219B}, would also result in plasma emission around 1-3\,GHz. Furthermore, \citet{kavanagh2021} predict that AU\,Mic could have a planet within the Alfvén surface, potentially allowing for radio emission to be produced through SPI at frequencies in the gigahertz regime.

Radio emission {in the gigahertz regime} has previously been detected on AU\,Mic. \citet{1985ASSL..116..233C} detected time-varying emission from AU\,Mic, which increases in brightness with frequency, up to 17\,GHz. \citet{kundu1987} detected a bright, strongly circularly polarised burst with the Very Large Array (VLA){ at 1.4\.GHz} that faded during their 3.5\,h observation, which they interpret to be likely driven by ECMI. However, this could not be confirmed as the frequency coverage was too small to detect structure. 

In this work, we present a year-long monitoring campaign of AU\,Mic using the Australia Telescope Compact Array (ATCA){ L-band receiver}, consisting of over 250\,h of data. Our observations have a large fractional bandwidth, between 1.1 and 3.1\,GHz, allowing us to analyse the time-frequency structure of AU\,Mic's radio emission in detail. The long time baseline of our observations allows us to perform a high-fidelity periodicity search over frequencies that cover the rotation rate of AU\,Mic and the orbital period of its innermost planets. Finally, with the long observing period and regular sampling, we can draw conclusions about the variation in radio emission on both short and long timescales. 

With this dataset, we aim to describe and catalogue the radio emission detected on AU\,Mic in a classification system that can encompass other radio stars as well. We also aim to search the data for periodicity, to confirm the emission mechanism and determine the location of origin of the emission.

In Section\,\ref{sec:obs}, we describe the data and methods used in this work. Section\,\ref{sec:phen} describes the detected radio emission, as well as a classification system based on the observed morphology of emission. Section\,\ref{sec:stats} describes the brightness distribution of the emission. In Section\,\ref{sec:structure}, we discuss the implications of the detected emission for the structure of the corona of the star. Section\,\ref{sec:period} covers the search for periodicity in the data, as well as the interpretation of the signal. Finally, in Section\,\ref{sec:variation}, we discuss possible long-term variations in the radio emission.

\section{Radio observations and data reduction}
\label{sec:obs}
We observed AU\,Mic regularly with the ATCA (Project IDs: C3267, CX513, and C3506, PI: Callingham; and C3469, PI: Climent) between March 2022 and March 2023, totalling over 250 hours. 
The observing setup and epochs are listed in Table\,\ref{tab:atca}. All observations were conducted {at 1.1-3.1\,GHz} using the 2\,GHz Compact Array Broadband Backend \citep[CABB;][]{Wilson2011}. The configuration of the array in the different epochs varies, with most observations conducted with the array in a standard 6-km configuration. All observations were taken using a 10\,second integration time, except for the epoch taken on \mbox{2022-07-17}, which was observed using a 3\,second integration time. All observations were taken with 1\,MHz channel widths, except the 2022-06-03 epoch, which had the correlator set up to have channel widths of 64\,MHz. Due to the poor radio-frequency interference (RFI) conditions {at these frequencies}, nearly all of the data of this epoch was flagged. As such, we exclude this epoch from our analysis but mention it for completeness. The other epoch that resulted in no reliable data was 2022-08-03, as it contained only 15 minutes on source at low elevation. The second half of the 2022-11-13 epoch also produced unusable data due to a nearby thunderstorm. During several of our epochs, one of the antennae of the ATCA was not included. For the 2022-09-15 epoch, antenna CA04 was offline. Antenna CA01 was not available on 2022-12-08. During the epochs in February and March 2023, antenna CA06 was not available. On 2023-03-08, antenna CA02 had to be disconnected soon after the start of the observation.

All of our data were reduced using {\ttfamily \selectfont CASA} \citep[v\,6.4,][]{2022PASP..134k4501C}. Due to the radio frequency interference (RFI) conditions at {1-3\,GHz}, on average 50-60\% of the data was flagged. The data was flagged using the {\ttfamily \selectfont tfcrop} algorithm, combined with manual flagging after visual inspection. A number of short baselines were flagged completely when the RFI was too strong to salvage them.

In all observations, we use PKS\,B1934-638 as the flux density and bandpass calibrator, and PKS\,2058-297 or PKS\,2032-350 as the phase calibrator. PKS\,2058-297 was used for the majority of observations, with PKS\,2032-350 only being used for the observations marked with $\dagger$ in Table\,\ref{tab:atca}. PKS\,B1934-638 was observed at the start and end of each observation, in scans lasting between 5 to 10\,minutes. The phase calibrator was observed for 5 minutes after every 30 minutes on source in observations conducted up to and including September 2022, and for 2.5 minutes after every 40 minutes on target during the epochs after September 2022. 

We first determined the gain, bandpass, polarisation and polarisation leakage solutions on the flux density calibrator, using a solution interval of 60\,s. After transferring these solutions to the phase calibrator, we use the {\ttfamily \selectfont qufromgain} task from the {\ttfamily \selectfont atca.polhelpers} package for linear polarisation calibration.
The combined solutions were then transferred to the target.
To create dynamic spectra in both Stokes\,I and Stokes\,V\footnote{We define the sign of Stokes\,V as right-handed circularly polarised emission minus left-handed circular polarised emission, in agreement with the IAU convention \citep{1996A&AS..117..161H}.} of our data, we first subtracted all Stokes\,I sources other than AU\,Mic. We did so by first imaging the data over 9600 by 9600 pixels (4\degree by 4\degree), using a pixel size of 1.5$''$ using \texttt{WSclean} \citep{2014MNRAS.444..606O}. The \texttt{WSClean} model of the field (not including AU\,Mic) was then subtracted from the calibrated visibilities using the CASA task {\tt uvsub}.
We then used {\texttt DP3} \citep{2018ascl.soft04003V} to phase-shift each measurement set to the exact measured location of AU\,Mic in each epoch, determined based on the Stokes\,I detection of AU\,Mic. Phase shifting ensured that AU\,Mic's emission only appeared in the real part of the visibility. From the phase-shifted measurement sets, we read in the residual visibility data using the {\tt pyrap} package\footnote{See \url{https://code.google.com/archive/p/pyrap/}}. 
The visibilities were then averaged across baselines which corresponds to natural weighting. The noise in the resulting dynamic spectrum was often too high to be able to resolve any structure, so we binned this data both in time and frequency. The averaging of the data was performed heuristically based on the signal-to-noise ratio to reliably resolve small-scale structures. While the real part of the averaged visibility in each bin gave us the flux density of AU\,Mic, the standard deviation of the imaginary part gave us the empirical measurement noise.

\begin{table}
\caption{\label{tab:atca} Summary of the observations used in this work. Epochs marked with an asterisk are epochs where the data was not salvageable. All epochs used PKS\,2058-297 as a phase calibrator, except the epochs marked with $\dagger$, which used PKS\,2032-350. The total amount of usable observing time is around 255 hours.} 
\begin{center}
\begin{tabular}{lrr}
\hline
Epoch  & Array configuration & Duration (hr)   \cr
\hline
2022-03-02 & 6A & 2 \cr
2022-03-06 & 6A & 3\cr
2022-04-17 & 1.5A & 10\cr
2022-04-18 & 1.5A & 10\cr
2022-05-29 & H214 & 5\cr
2022-06-23$^*$ & H214 & 4\cr
2022-07-17 & H214 & 5\cr
2022-07-25 & H214 & 7 \cr
2022-07-31 & H168 & 7.5\cr
2022-08-03$^*$ & H168 & 0.5\cr
2022-08-18 & H168 & 5\cr
2022-08-19 & H168 & 5 \cr
2022-09-09$^{\dagger}$ & 6D & 6 \cr
2022-09-09 & 6D & 3.5 \cr
2022-09-10$^{\dagger}$ & 6D & 6 \cr
2022-09-11$^{\dagger}$ & 6D & 6 \cr
2022-09-12$^{\dagger}$ & 6D & 6 \cr
2022-09-13$^{\dagger}$ & 6D & 6 \cr
2022-09-14$^{\dagger}$ & 6D & 6 \cr
2022-09-15$^{\dagger}$ & 6D & 6 \cr
2022-09-15 & 6D & 6\cr
2022-09-16$^{\dagger}$ & 6D & 6 \cr
2022-09-17$^{\dagger}$ & 6D & 6 \cr
2022-10-02 & 6D & 6\cr
2022-10-07 & 6D & 4\cr
2022-10-24 & H214 & 5\cr
2022-10-27 & H214 & 8\cr
2022-11-06 & EW352 & 9.5\cr
2022-11-13 & 6C & 4\cr
2022-11-23 & 6C & 5\cr
2022-11-27 & 6C & 4\cr
2022-12-08 & 6C & 4\cr
2022-12-09 & 6C & 4\cr
2022-12-25 & 6C & 9.5\cr
2022-12-31 & 6C & 10\cr
2023-01-19 & 6C & 8.5\cr
2023-02-02 & 6C & 9\cr
2023-02-04 & 6C & 6\cr
2023-02-26 & 1.5B & 10.5\cr
2023-03-01 & 750C & 8\cr
2023-03-03 & 750C & 10.5 \cr
2023-03-08 & 750C & 4 \cr
2023-03-18 & 750C & 6 \cr
2023-03-31 & 750C & 6.5 \cr

\hline
\end{tabular}
\end{center}
\end{table}

\subsection*{Technical challenges}
The majority of our observations were taken in a non-hybrid array, meaning all baselines lie in the east-west direction. This results in some unique challenges when discerning small timescale variations.
When the array is in an east-west configuration, the instantaneous point-spread function (PSF) is narrow along the east-west axis, but extremely long along the north-south axis. This one-dimensional PSF produces an insidious sidelobe contamination from nearby sources in the dynamic spectra. 

In the dynamic spectra, sidelobes of nearby sources appear as arcs in time-frequency space. The shape is caused by the sidelobes of these sources changing as a function of frequency, combined with the rotation of the sky, leading to the sidelobes sweeping across the phase centre. In most cases, this does not affect our interpretation, as these arcs are easy to recognise. However, when the one-dimensional instantaneous PSF of a source, either in the primary beam or in the sidelobes of the beam, lines up exactly with the phase centre, it may produce a structure not dissimilar to a solar Type\,II burst \citep{dulk1985}, lasting for a few minutes with a tentative sweep structure. The flux density of this structure will be the flux density of the original source without primary-beam correction. We largely mitigated this effect by subtracting the model of all sources in the field, but this did not fully remove the patterns in time-frequency space. The patterns have a predictable structure in the dynamic spectra and can readily be excluded from analysis. Nevertheless, we caution the reader that narrow time-frequency structures in Stokes\,I dynamic spectra must be treated with caution. 

Stokes\,V dynamic spectra are not affected by sidelobes of nearby sources, as there are no other bright circularly polarised sources within our field of view. Polarised emission can therefore be reliably identified. Small-scale variations in time and frequency of unpolarised emission is more difficult to identify. In cases where the Stokes\,I dynamic spectrum shows interesting structure, we re-calibrated the data as carefully as possible to decrease the effect of sidelobes. Even after this process, Stokes\,I time-frequency structure should be interpreted with caution.

\section{Phenomenology of radio emission}
\label{sec:phen}
\begin{figure*}
    \centering
    \begin{subfigure}[t]{0.45\textwidth}
    \captionsetup{width=.9\textwidth}
    \includegraphics[width=\columnwidth]{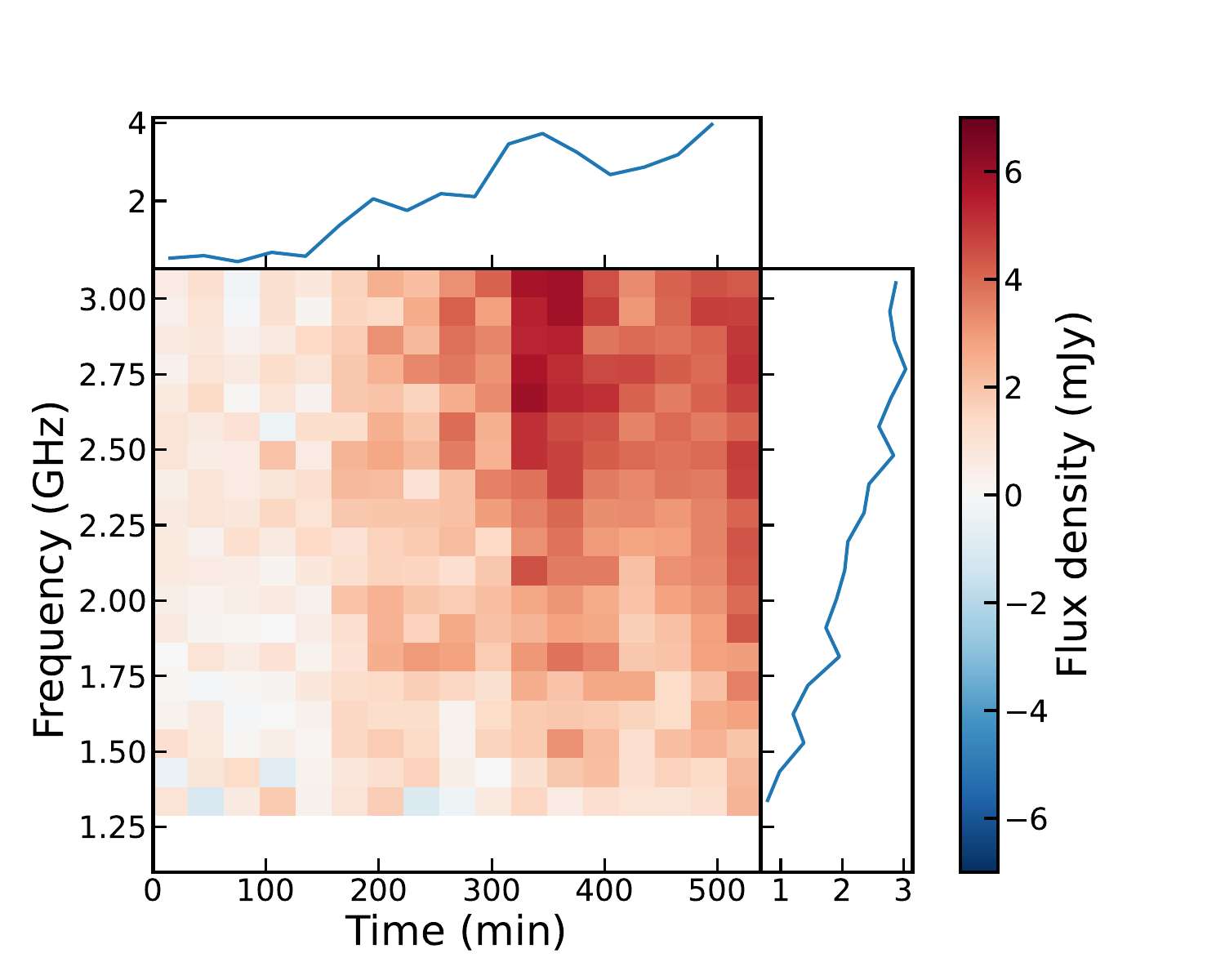}
    \caption{Type\,A, broadband emission (Stokes\,I).}
    \label{fig:gyro}
    \end{subfigure}
    \begin{subfigure}[t]{0.45\textwidth}
            \centering
            \captionsetup{width=.9\textwidth}
    \includegraphics[width=\columnwidth]{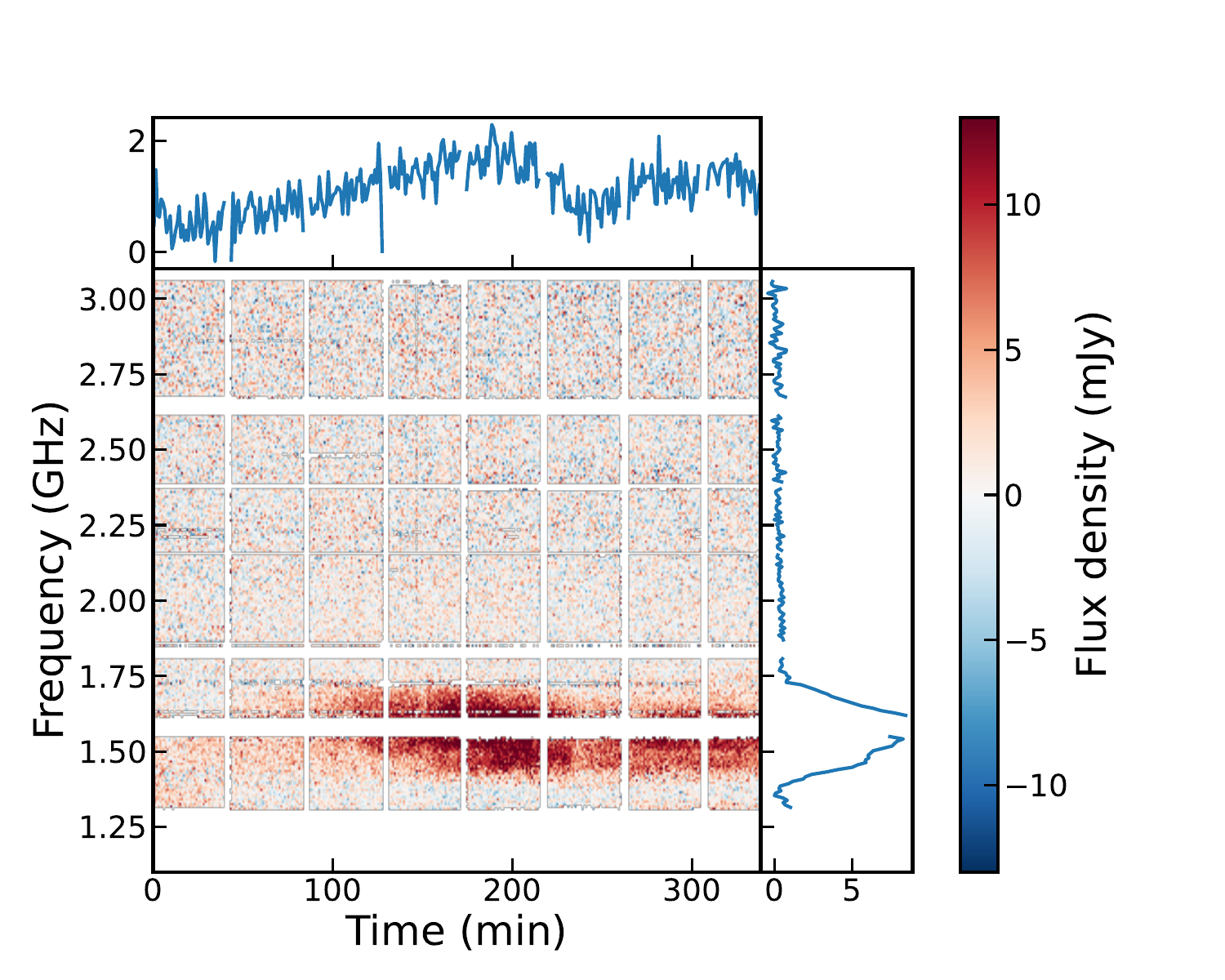}
    \caption{Type\,B, a confined burst (Stokes\,V).}
    \label{fig:confined}
    \end{subfigure}
\begin{subfigure}[t]{0.45\textwidth}
    \centering
    \captionsetup{width=.9\textwidth}
    \includegraphics[width=\columnwidth]{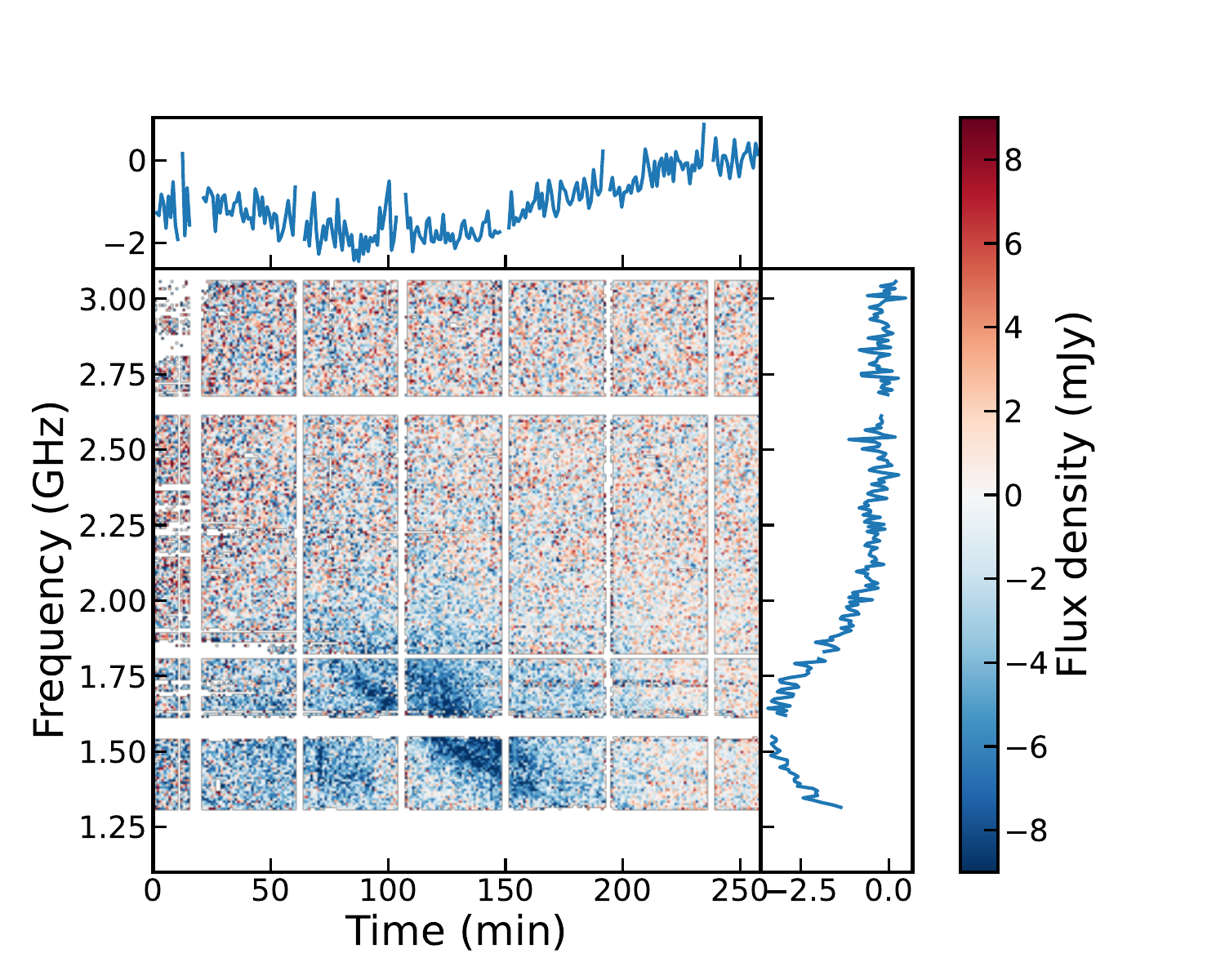}
    \caption{Type\,C, a slow sweep burst (Stokes\,V).}
    \label{fig:2022-11-23}
\end{subfigure}    
\begin{subfigure}[t]{0.45\textwidth}
    \centering
    \captionsetup{width=.9\textwidth}
    \includegraphics[width=\columnwidth]{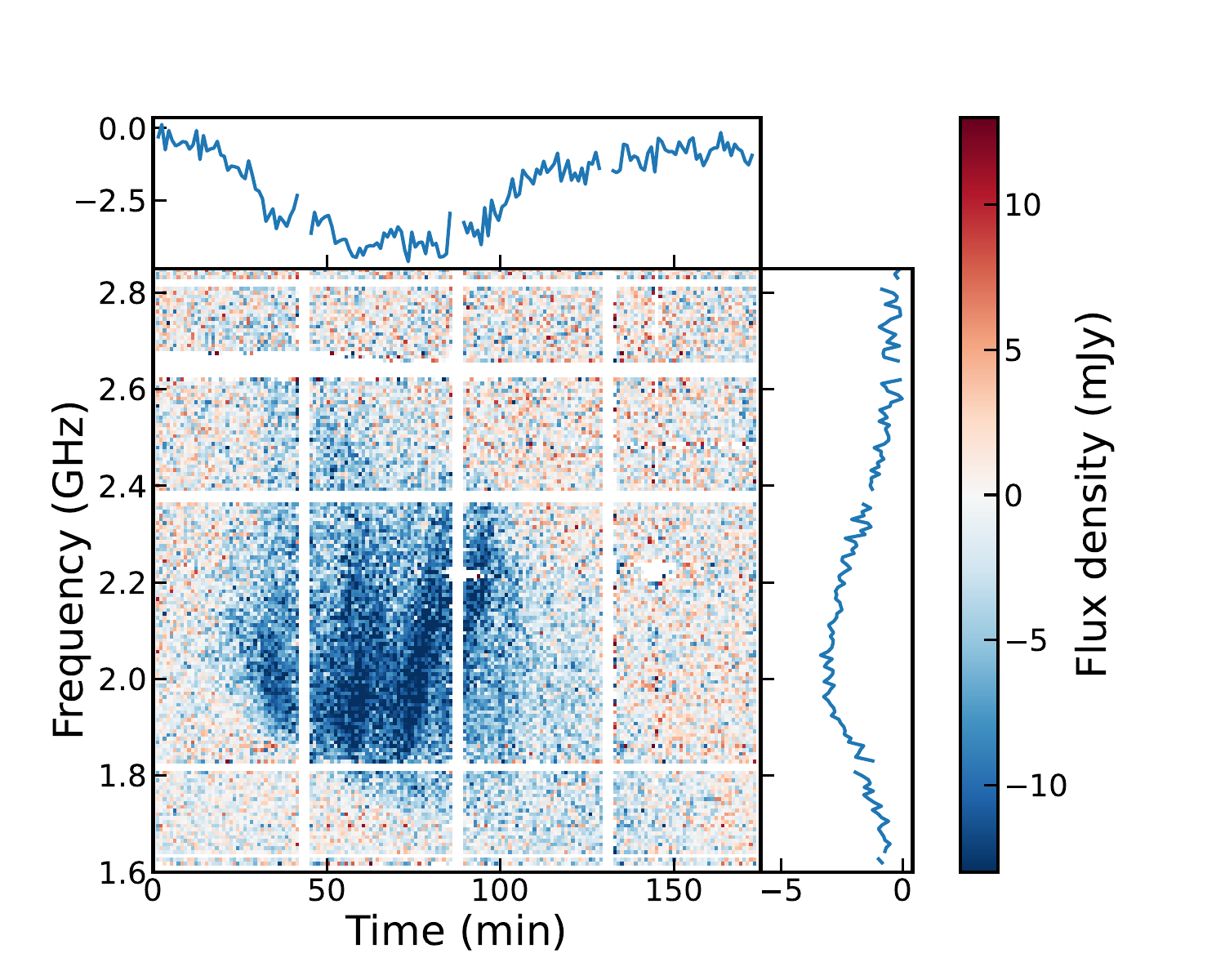}
    \caption{Type\,D, a fast sweep burst (Stokes\,V).}
    \label{fig:2022-11-27}
\end{subfigure}

\begin{subfigure}[t]{0.45\textwidth}
    \centering
    \captionsetup{width=.9\textwidth}
    \includegraphics[width=\columnwidth]{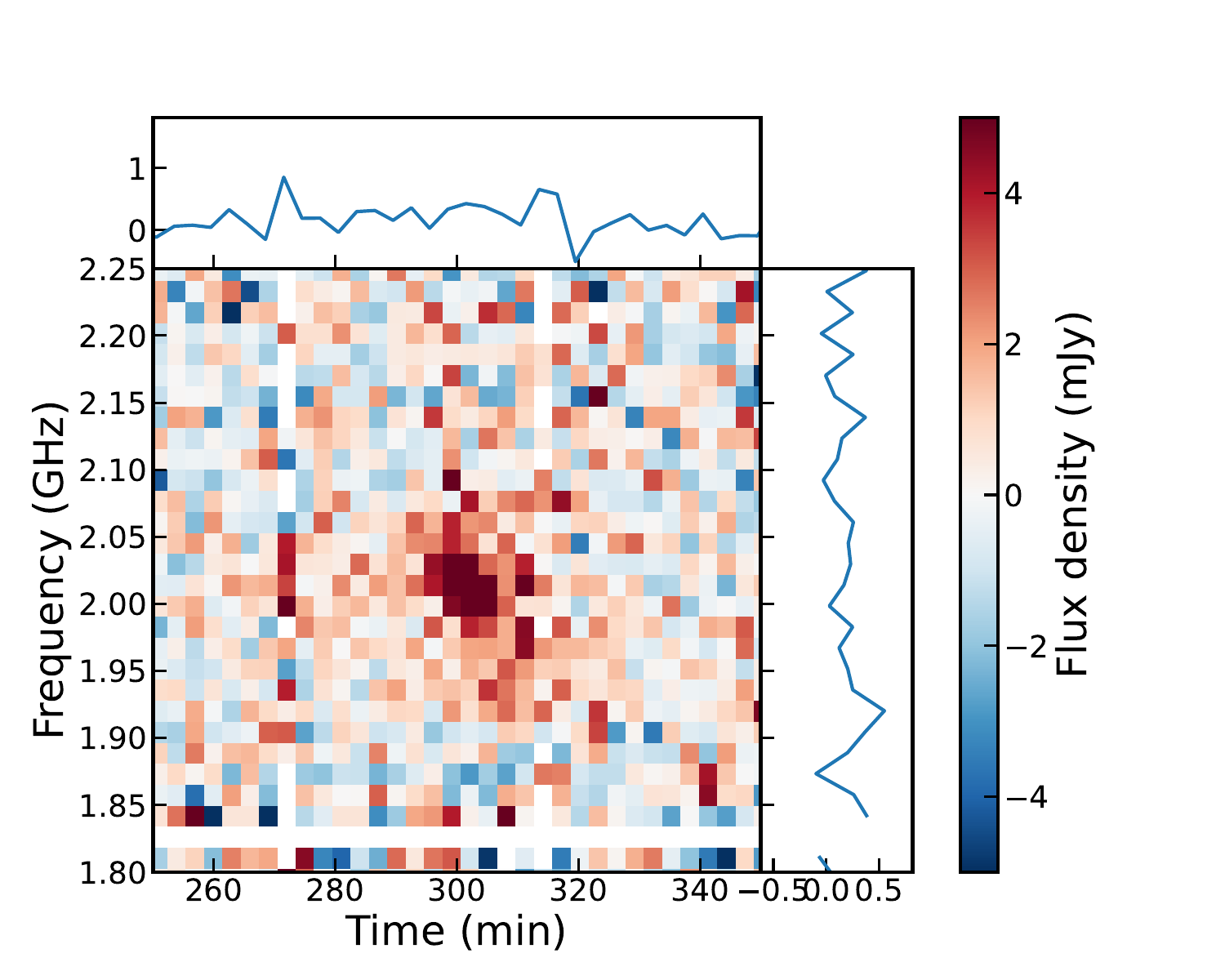}
    \caption{Type\,E, a shot burst (Stokes\,V).}
    \label{fig:small}
\end{subfigure}
\begin{subfigure}[t]{0.45\textwidth}
    \centering
    \captionsetup{width=.9\textwidth}
    \includegraphics[width=\columnwidth]{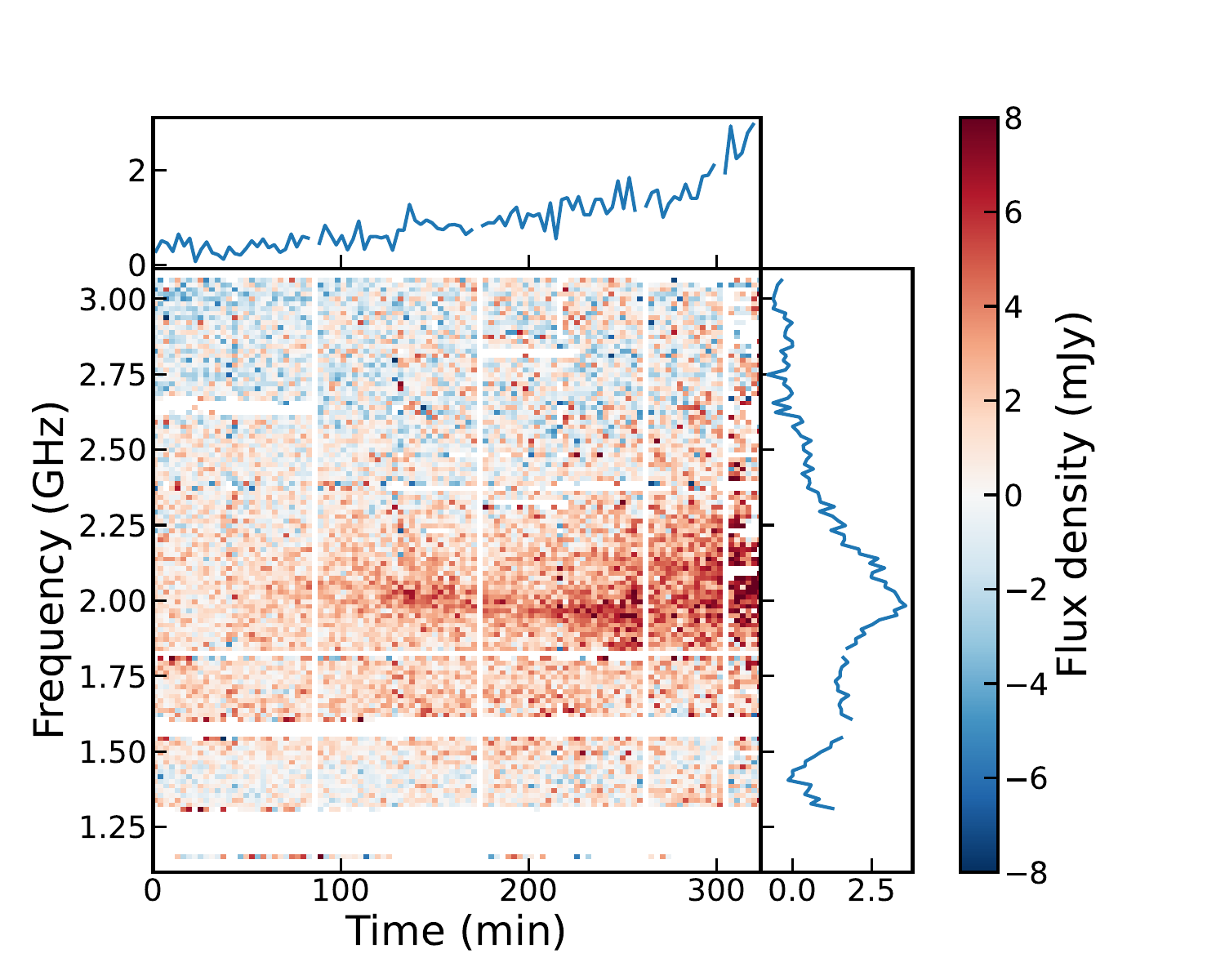}
    \caption{Type\,F, an irregular burst (Stokes\,V).}
    \label{fig:unconfinec}
\end{subfigure}
\caption{Dynamic spectra of the 6 different types of emission detected on AU\,Mic. Note that the colour scale and the axes are not the same for each figure. Panel\,(a) is binned into time bins of 30 minutes and frequency bins of 100\,MHz. Panels\,(b) through (e) are binned into time bins of 1 minute and frequency bins of 8\,MHz, and Panel\,(f) is binned into time bins of 3 minute and frequency bins of 16\,MHz. For each plot, the colour bar shows the flux density, with frequency on the vertical axis and time on the horizontal axis. On the top, a light curve is plotted, averaged over all frequencies. To the side, we plot a spectrum of the emission, averaging over all time integrations. White regions in the plots indicate parts of the data that have been completely flagged.}
\label{fig:categories}
\end{figure*}

As this is the longest radio monitoring campaign on a star outside the Solar System, we can search for patterns in the time-frequency structure of the emission.
We detect radio emission on AU\,Mic in 40 out of our 42 epochs. The type of emission we detect varies significantly. In this section, we highlight the different categories of radio emission detected in our survey, and present a classification scheme based on the observed properties. The classification scheme is purely phenomenological, based on the observed time-frequency structures and polarisation properties of the emission. Although it partially overlaps with the solar radio burst classification, this system is designed to apply to a wider variety of stars, in particular those with strong magnetic fields. We demonstrate how this scheme can be applied to radio emission detected from other M\,dwarfs in the literature. {Table\,\ref{tab:types} shows the descriptions of the different types of emission, including their classification. We also list the likely emission mechanism. Examples of each category detected on AU\,Mic are shown in Fig.\,\ref{fig:categories}.}

\begin{table*}
\resizebox{\linewidth}{!}{
    \centering
    \begin{tabular}{llllllll}
    \hline \hline
Classification & Type & (N, M) & Timescale & Bandwidth & |V|/I & Internal structure  & Interpretation\cr
&  & & & (Fractional width) & & \cr
\hline
Type A & Broadband emission & (40,42)& >3 h & >2\,GHz & <30\% & Smooth emission in time and frequency. & Incoherent \cr
& & & &(>1) & & The spectrum has a positive spectral index. & {(Gyromagnetic)} \cr
\cr
Type B & Confined burst & (6,42) & 3-6 h & 200-500\,MHz & >90\% & Bursts that are strongly confined in frequency. & Coherent\cr
& & && (0.1-0.5) & & {Do not show prototypical internal frequency structure.} & (ECMI)\cr
\cr

Type C & Slow sweeps & (9,42)& 1-3\,h & 200-500\,MHz & >90\% & These bursts move to lower or higher frequencies over time. & Coherent\cr 
&&& &(0.07-0.5) && {Do not change direction in frequency and can have multiple lanes of emission.}& (ECMI) \cr
&&& & && The drift rate is 5\,MHz/min or lower.\cr
Type D & Fast sweeps & (1,42)& 0.5-1\,h & 400-700\,MHz & >90\% & Bursts with larger drift rates around 50\,MHz/min. & Coherent\cr
&&&& (0.13-0.6)&& They can have several sweeps in a short time, & (ECMI)\cr&&&&&& which might not move in the same direction. \cr
Type E & Shot bursts & (2,42)& <15\,min & < 100\,MHz & >50\% & {These bursts look like bullet holes in the dynamic spectrum.} & Coherent\cr
&&&&(<0.025)&& {They are strongly confined in both frequency and time.} & (ECMI?) \cr
&&&&&& {No detected internal time-frequency structure.} & \cr
Type F & Irregular bursts & (4,42)& 1-6\,h & 0.3-1.5\,GHz & >90\% & Similar to the confined and sweep bursts, but less constrained. & Coherent\cr
&&&&(0.1-1)&& Moves around in frequency and can change directions. & (ECMI)\cr
\cr
Type G & Pulse bursts & --- & 5\,min-1\,h & 60\,MHz-6\,GHz & >90\% & These bursts are broadband and often last a few minutes. & Coherent\cr
&&&&(>0.5)&& They can have strong variations on short timescales. & (---) \cr
&&&&&&Not detected on AU\,Mic.\cr

\hline \hline
    \end{tabular}}
    \caption{Summary of our classification system for the different types of radio emission detected on stars. Examples of each category detected on AU\,Mic are shown in Fig.\,\ref{fig:categories}. Several bursts fit into more than one category, in which case they are counted in both. There can also be multiple separate types of emission in one epoch. (N, M) indicates the number of epochs in which emission of that type has been detected, compared to the total number of epochs. |V|/I shows the absolute value of the circular polarisation fraction. The last column shows whether this type of emission is coherent or incoherent, along with our interpretation of this type of emission when detected on AU\,Mic.}
    \label{tab:types}
\end{table*}

{
The Type\,A broadband emission shown in Fig.\,\ref{fig:gyro} increases in brightness around halfway through the observation. We only detect such a strong variation once in 40 detections of Type\,A emission. Although Type\,A emission is in general not strongly variable on a timescale of hours, the flux density of this emission varies by over an order of magnitude between observations.
}
{
The handedness of polarisation of the types of emission varies. Most types of emission can produce both left-handed and right-handed emission. Type\,E bursts seem to be an exception, as both detections of this type of emission show right-handed circular polarisation. However, the number of detections is too small to draw any conclusions.}

\subsection*{Classifying literature detections of stellar radio bursts}
With the classification system we have developed, we can go back to previously published work and determine whether it can {potentially} be universally applied to other radio bright stars.
Previous radio detections of AU\,Mic by \citet{kundu1987} showed a coherent burst on a timescale of hours. The data was taken with the VLA before the broadband back-end was installed, so the frequency structure of the burst remains unclear. Based on the timescale, we suggest that this burst most likely fits into Type\,B, Type\,C or Type\,F. Another radio detection of AU\,Mic \citep{1985ASSL..116..233C} shows a broadband emission pattern {between 1 and 15\,GHz}, resembling the Type\,A emission presented in this work. 

Several literature examples of detections of bursts from other M\,dwarfs are also available. For example, \citet{villadsen2019} detected 22 coherent bursts from a sample of M\,dwarfs -- a study very much complementary to our own, except covering a sample of M\,dwarfs, instead of a single star for hundreds of hours. Based on the published dynamic spectra of these stars, we would classify the emission in their 13 epochs as five to six Type\,F bursts, three Type\,B bursts (on YZ Canis Minoris and AD\,Leonis), two bursts that are classified as Type\,B and C, one Type\,D burst, and eight to nine epochs with Type\,A emission. 

A few of the bursts observed by \citet{villadsen2019} do not fit neatly into our categories. These bursts are very broadband and short compared to the structures we see on AU\,Mic. These could be even faster sweep bursts than those detected on AU\,Mic. We note that most of these bursts are detected on UV\,Ceti, which is an M6\,dwarf, the latest spectral type in their sample. Other works on UV\,Ceti, such as \citet{zic2019} and \citet{Bastian_2022}, show similar structures of bursts sweeping extremely quickly in frequency and lasting for $\sim$10 minutes. 
To make our phenomenological classification scheme more universally applicable, we added Type\,G to our classification system. Type\,G bursts include broadband bursts that can last for a few minutes up to an hour. The prototypical examples for this category are two bursts observed by \citet{villadsen2019} on UV\,Ceti, on 2013-05-26.

On YZ Ceti, \citet{pineda2023} detected two coherent bursts and broadband emission. The broadband emission fits into Type\,A, whereas the two coherent bursts fit into Type\,G.
\citet{2006ApJ...637.1016O} and \citet{2008ApJ...674.1078O} observed a number of radio bursts on AD\,Leonis, which we classify as Type\,G and one Type\,F. On EQ Pegasi, \citet{2018ApJ...862..113C} detect a burst that we classify as Type\,E. At lower frequencies, \citet{callingham2021} detected several radio bursts on CR\,Draconis. Of these bursts, we classify one as a Type\,C burst and two as a Type\,G burst.

In conclusion, all types of bursts that we detect on AU\,Mic also occur on other M\,dwarfs. One category of bursts in our classification scheme is not detected on AU\,Mic, but is detected on several other M\,dwarfs. Our classification scheme can be of use to future studies of radio stellar systems as a way to compare the multi-dimensional time-frequency structure discussion that occurs with high-resolution dynamic spectra.

\section{Brightness distribution of the bursts}
\label{sec:stats}
As stated in Section\,\ref{sec:phen}, AU\,Mic produces strongly variable radio emission. In this section, we present the brightness distribution of the radio bursts. Such a distribution has not been done in the radio regime for a single star, but can be critical in determining whether low or high energy flares are heating the corona of the star \citep{Kashyap_2002}.

We exclude the Type\,A broadband emission in calculating the distribution. We only consider clear bursts because Type\,A emission often occurs in combination with a burst, and disentangling the peak fluxes of the two different types of emission is difficult. Therefore, we focus on only the Stokes\,V emission from coherent bursts, as it is the most reliable and the least affected by Type\,A emission. For this analysis, we count bursts as separate if they are not overlapping in time-frequency space.

The Stokes\,V flux densities used here are determined by binning the emission to 15\,minutes and 50\,MHz. For each burst, we use the flux density of the bin with the highest signal-to-noise value, where the error is determined by taking the standard deviation of the imaginary component of all data points in each bin.

\begin{figure}
    \centering
    \includegraphics[width=\columnwidth]{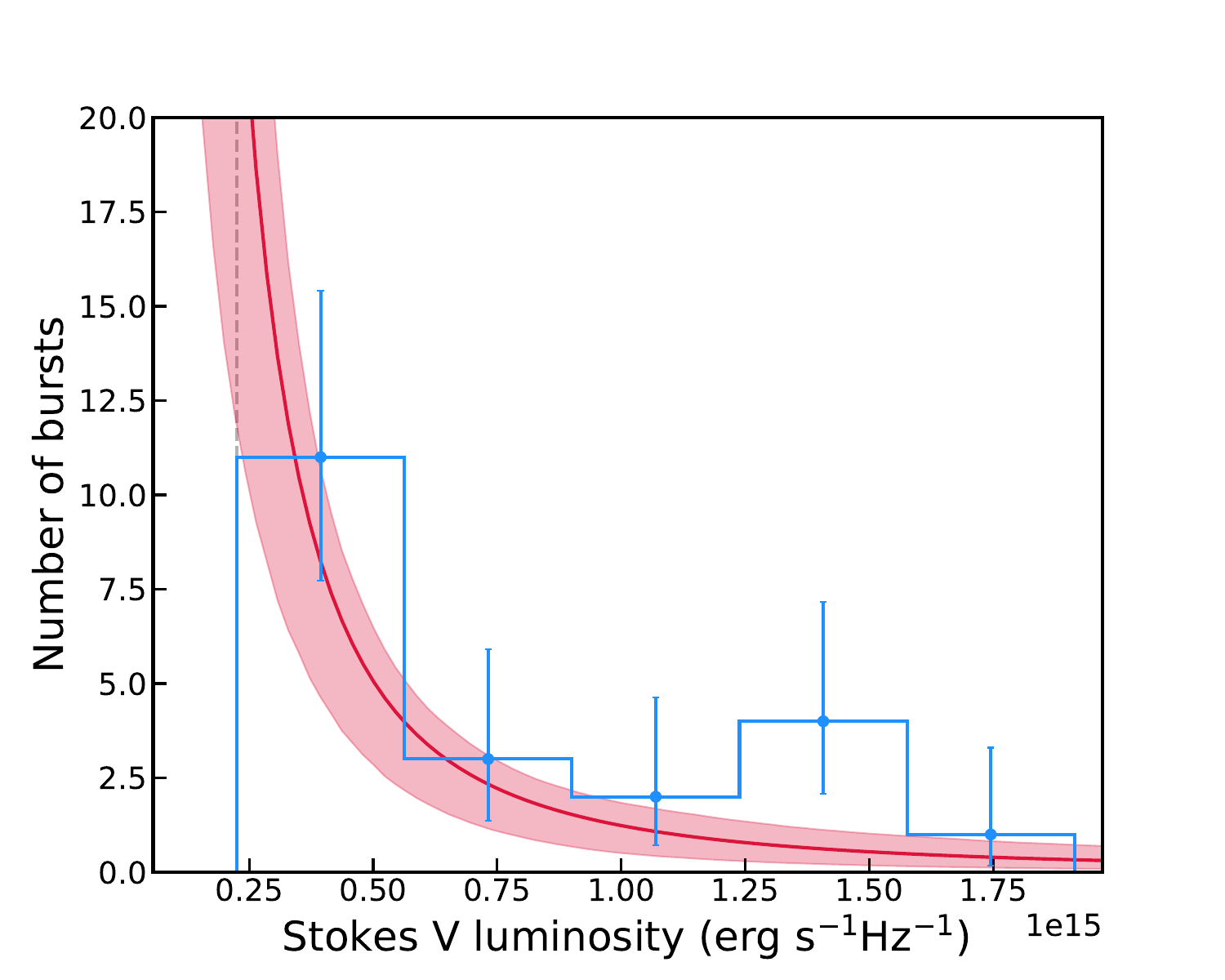}
    \caption{Distribution of the luminosity in Stokes\,V for all bursts with a peak flux density over 2.0\,mJy. The grey dashed line shows this cutoff of 2.0\,mJy, used to select only burst emission. The red line shows the best fit to the histogram, with the shaded region showing the 1$\sigma$ uncertainties.}
    \label{fig:hist}
\end{figure}

As our sensitivity varies slightly between observations due to different numbers of antennae and variable RFI conditions, the exact level of completeness is unclear. To be conservative, we only include bursts with a peak flux density in Stokes\,V over 2\,mJy.
Figure\,\ref{fig:hist} shows the distribution of luminosities of all bursts that satisfy this requirement. We note that the luminosity was calculated assuming isotropic emission. The emission could be beamed as well, in which case the luminosities given here are overestimated. The error bars are determined using a prescription for low-number statistics by \citet{1986ApJ...303..336G}. The number of bursts above our threshold is high considering previous radio studies of stars, but low when compared to studies of flares at other wavelengths \citep[e.g.][]{2020AJ....159...60G, gilbert2022}. 

While we are impacted by small number statistics, the distribution clearly shows a higher number of bursts at a low flux density when compared to the brightest bursts. 
We fitted a power law model to the data, described by  \begin{equation}
    N(S)=N_{0} \left(\frac{S}{2}\right)^{\beta},
\end{equation}
where $N_0$ is the normalisation constant, $S$ is the flux density, and $\beta$ is the index of the power law. The best-fit result is shown in Fig.\,\ref{fig:hist}, with $N_0=25\pm13$ and $\beta=-2.0\pm0.75$. The uncertainties on these values are too large to test whether low or high energy flares play a dominant role in coronal heating. However, this is the first measurement of burst radio luminosity distribution on a star other than the Sun.

Extrapolating the best-fit relationship to lower flux density values, we can determine how many bursts would be observed with a more sensitive instrument. The noise of MeerKAT at a similar time-frequency resolution is approximately 0.12\,mJy. Setting the detection threshold at 5$\sigma$, we find that MeerKAT should detect around 2.5$\pm$1 times as many bursts as ATCA. However, as discussed in more detail in Section\,\ref{sec:period}, the dominant factor that determines whether a burst is observed is most likely not the sensitivity of the instrument.

\section{Emission mechanism}
\label{sec:structure}
With the large amount of radio data and the categorisation of emission, we now detail the likely emission mechanism behind each observed type. Here, we split the observed emission into two overarching categories: radio bursts, consisting of all coherent emission, and broadband emission, consisting of only Type\,A emission.

\subsection{Radio bursts}
\label{sec:em}
The observed radio burst structures can be produced in several ways. The bursts are highly circularly polarised, showing that they must be produced through a coherent emission mechanism. The possible coherent emission mechanisms are plasma emission and the electron cyclotron maser instability (ECMI).

\subsubsection{Plasma emission}
First, we consider plasma emission. The frequency at which this emission is produced is given by 
\begin{equation}
    \label{plasma_freq}
    \nu_{\mathrm{p}} \approx 9000 \,\left(\frac{n_\mathrm{e}}{{\rm cm}^{-3}}\right)^{1/2}\,\,{\rm Hz},
\end{equation}
and its harmonics, where $\nu_\mathrm{p}$ is the plasma frequency and $n_\mathrm{e}$ is the local electron density \citep{dulk1985}.
Due to the high observed circular polarisation fraction that is often close to 100\%, the emission would have to be fundamental plasma emission. The emission can be produced if there is a population of high-energy electrons that causes a bump-in-tail instability, which generates Langmuir waves.
On the Sun, plasma emission often produces short bursts that last for a few seconds to a few minutes, with a comparatively low luminosity. The bursts we see on AU\,Mic are generally on a timescale of hours, and very luminous. The only type of solar burst on a similar timescale and energy level is the moving Type\,IV burst. These bursts are thought to be caused by a moving plasmoid that is slowly expanding in the Solar corona, leading to a slow drift to lower frequencies, similar to the Type\,C emission described in Section\,\ref{sec:phen}.

The handedness of the circularly polarised component of plasma emission depends on the direction of the magnetic field in the region of emission. Radiation can be produced in ordinary (o-mode) or extraordinary mode (x-mode). These modes are defined relative to the direction of the magnetic field at the location of emission. 
While plasma emission can be produced both in o- and x-mode, x-mode emission can only propagate at a frequency above the x-mode cutoff \citep{2002ASSL..279.....B}, given by
\begin{equation}
    \nu_{\mathrm{x}} = \sqrt{\nu_{\mathrm{p}}^2 + \frac{1}{4}\nu_{\mathrm{B}}^2} + \frac{1}{2}\nu_{\mathrm{B}},
\end{equation}
where $\nu_{\mathrm{B}}$ is the cyclotron frequency, given by \begin{equation}
    \label{eq:cyc_freq}
    \nu_{\mathrm{B}} \approx 2.8 \times 10^6 \left(\frac{B}{\rm Gauss}\right)\,\,{\rm Hz}.
\end{equation}
At the fundamental, the emission is therefore close to 100\% circularly polarised in the sense of the o-mode.

Assuming that this emission is fundamental plasma emission, the frequency where the emission is produced results in a local plasma density between 1.5$\times 10^{10}$\,cm$^{-3}$ and 1.2$\times 10^{11}$\,cm$^{-3}$. These numbers are in agreement with measurements of the plasma density on AU\,Mic \citep[e.g.][]{2002A&A...390..219B}, although it is unclear whether these measurements truly represent the base plasma density, or if they represent local variations in density, for example those caused by flares.
The brightness temperatures of the bursts range from $3\times 10^9$\,K to $6\times 10^{10}$\,K if we assume the source covers the entire stellar disk. This range can be fully explained by fundamental plasma emission, which has a brightness temperature limit at $10^{15}$\,K \citep{dulk1985}.

\subsubsection{Electron Cyclotron Maser Instability}
The other coherent emission mechanism, ECMI, also produces highly circularly polarised bursts. The requirement for ECMI emission to be produced is a population of high-energy electrons with a population inversion, often assumed to be a loss-cone or a horseshoe distribution \citep{2006A&ARv..13..229T}. This emission is produced at the harmonics of the cyclotron frequency, $\nu_{\rm B}$, which, in the non-relativistic limit, only depends on the ambient magnetic field strength, $B$, as shown in Eq.\,\ref{eq:cyc_freq}. ECMI can be produced in both the large-scale field, as seen on Jupiter \citep[e.g.]{marques2017}, and in small-scale loops as seen on the Sun \citep[e.g.]{sunspot}.
If the emission is driven by ECMI, the frequencies at which we have detected bursts from AU\,Mic imply the star possesses magnetic field strengths between 390\,G and 1.1\,kG. These values are in agreement with the measured maximum surface magnetic field strength of 1-2\,kG from Zeeman-Doppler imaging (ZDI) maps \citep{2020ApJ...902...43K,zdi_aumic}.
Assuming a maximum dipole surface strength of 2\,kG, ECMI emission at the fundamental can be produced at most 0.8 stellar radii away from the surface. If the emission is at the second harmonic of the cyclotron frequency, it can be produced at most 1.2 stellar radii from the surface.

ECMI can produce high brightness temperatures, having a higher brightness temperature limit than plasma emission.
The brightness temperatures of the bursts are consistent with both ECMI and plasma emission. However, unlike plasma emission, ECMI emission is strongly beamed as it is emitted along the thin surface of a cone with a wide opening angle and axis aligned with the ambient magnetic field vector \citep{1982ApJ...259..844M}. This, when combined with rotation, results in burst structures in the pattern of emission. On Jupiter, which also produces radio emission through ECMI, the bursts often look like arcs that last for a few hours, with the direction of the frequency sweep depending on the side of the cone we are observing \citep{marques2017}. The frequency structure of the emission is determined by how the cone width varies with frequency. As a function of time, ECMI bursts can move to higher and lower frequencies, or remain at the same frequency, depending on the beaming configuration of the system.
Since this emission is strongly beamed, it is often only visible for a fraction of the time.

The handedness of the circular polarisation of ECMI also depends on the local direction of the magnetic field, similar to plasma emission. The expected mode of emission depends on the local plasma density. While both o- and x-mode can be produced, the growth rate of x-mode dominates when $\nu_{\mathrm{p}}/\nu_{\mathrm{B}} \lessapprox0.05$ \citep{1982AuJPh..35..447H}. At higher densities, the o-mode growth rate dominates at the fundamental harmonic, up to $\nu_{\mathrm{p}}/\nu_{\mathrm{B}} > 1.0$. At the second harmonic, the x-mode dominates.

{ECMI is thought to be strongly absorbed at the harmonics of the cyclotron frequency \citep[e.g.][]{1982ApJ...259..844M}, which would make escape from the corona highly unlikely. However, ECMI has been detected on several occasions \citep[e.g.]{zic2019,callingham2021, Bastian_2022}, showing that the emission can escape. While some explanations have been offered, the escape of ECMI from stellar magnetospheres is still an unsolved question.}

Based on only the observed time-frequency structure and brightness temperature, it is not possible to uniquely determine the emission mechanism driving the bursts. To confirm that the emission is driven by ECMI, we need to know if the emission is beamed. This will be discussed in more detail in Section\,\ref{sec:period}.

\subsection{Broadband component -- gyromagnetic emission}
In most epochs, we detect unpolarised or mildly circularly polarised broadband emission, which we classified as Type\,A. The broadband emission is very different in time-frequency structure from the bursts. It is smooth in both time and frequency, with no sudden variations or extreme cut-offs. The circular polarisation fraction is also much lower, being at most around 30\%. Although such characteristics imply an incoherent emission mechanism is responsible for generating the emission, it is possible to create such polarisation fractions with a coherent mechanism as well, when averaging over many emission sites across the entire surface of the star. However, since the broadband emission is smooth, broadband, and of a longer duration than the bursts, we think it is unlikely the emission is caused by either plasma emission or ECMI. We can also rule out that the detected emission is thermal emission. Although thermal emission can produce a spectrum similar to what we observe on AU\,Mic, it cannot reproduce the brightness temperatures of the detected emission \citep{dulk1985}, which range from {$3.6\times 10^8$\,K to $7.25\times 10^{9}$\,K, assuming the source of the emission covers the entire stellar disk and the radius of the star is 0.75 solar radii \citep{aumic_b}. The source of emission may be larger, as the emission may be produced further away from the star. This would result in lower brightness temperatures compared to those listed here.}

Having ruled out other emission mechanisms, we determine that the most probable emission mechanism of this broadband emission is {incoherent gyromagnetic} emission, which produces broadband emission that is roughly constant over time. {The frequency at which the emission peaks is determined by the strength of the magnetic field in the emitting region, momentum distribution of charges, and the angle between the emission and the magnetic field \citep{Rybicki_Lightman}. For commonly encountered momentum distributions, at frequencies above the peak, the emission is generally expected to look like a power law with a negative spectral index}, whereas the emission at frequencies below the peak is self-absorbed, leading to a positive spectral index with a limit of +2.5 \citep{Rybicki_Lightman}. The emission can be mildly circularly polarised, in the sense of the o-mode below the turnover frequency and in the sense of the x-mode above the turnover frequency.

The spectrum and brightness of {gyromagnetic} emission can vary over time, as it depends on the underlying electron distribution. {This agrees with our data, as the brightness of the broadband emission detected on AU\,Mic varies per epoch by over an order of magnitude.}

In our observations, we see some variety in the frequency structures of the quiescent, constant background emission. In some epochs, the spectral index of emission is positive and close to flat. In general, the spectral index of the emission is between 0 and 1. We do not see a clear turnover in the data, which implies that either the peak frequency is not within our observing band, or that we are seeing multiple populations of electrons producing the emission, leading to a number of peaks that are not distinguishable with our signal-to-noise ratio. 

{
Fitting the radio spectrum requires radiative transfer calculations with a realistic spatial distribution of charges and magnetic field strength. We leave this for future work. Instead, we check if approximate fitting formulae for gyromagnetic emission yield an energetically feasible accelerated charge population. We apply Eq.\,18 from \citet{Gudel2002}, assuming a magnetic field of 1\,kG, a peak frequency of 3\,GHz, a scale length equal to the radius of the star, and $\delta=2$, to determine the density of electrons with energies above 10\,keV. We find a density of the order of $10^3$cm$^{-3}$.
These particles constantly lose energy through Coulomb collisions, following Eq.\,2 from \citet{1979ApJ...227.1072B}. 
For a thermal plasma density of $10^9$\,cm$^{-3}$, we find that the energy supply rate must be $\sim3\times 10^{25}$\,erg\,s$^{-1}$ to compensate for these losses. The soft X-ray luminosity of AU\,Mic is $\sim 10^{30}$\,erg\,s$^{-1}$ \citep{1990A&A...228..403P}, so there is sufficient energy available in the corona to sustain the radio-emitting electron population.}

\section{Periodicity analysis} 
\label{sec:period}
As stated in Section\,\ref{sec:intro}, a star can produce radio emission by itself or through an interaction with a companion. When the star produces radio emission without a companion, it can be periodic with the stellar rotation or completely random in phase, depending on the type of emission. If the emission is produced through an interaction with a companion, the emission is expected to be periodic with the orbital period of the companion or the synodic period of the companion and the star, although this depends on the geometry of the system \citep{kavanagh2023}.
A detection of periodicity can therefore help to distinguish emission mechanisms and their possible causes. In this section, we create periodograms of the light curve of AU\,Mic to search for such a periodicity.

\subsection{Phase coverage}
Our observations are spread over 13 months, with epochs distributed fairly sparsely, resulting in a few epochs per month. The exception to this is the section of the observing campaign in September 2022, where we observed AU\,Mic for at least 6 hours every day for 9 days. Looking at the rotation period of the star and the orbital periods of the planets in the system, all periods are sampled fairly well, with no large gaps in the phase coverage. 

In what follows, with each periodogram of the data, we also present the periodogram of the window function for that data set. In general, we find no significant peaks in the window function that have strong associated components in the periodogram of the data. The strongest peak in the window function is located at a period of 23 hours and 56 minutes, or exactly one sidereal day. This peak is caused by the observing constraint that we can only observe AU\,Mic when it is above the horizon in Narrabri. The peak is strong because the observing campaign lasted for over a year, making the effect clearly noticeable.

\subsection{Lomb-Scargle periodogram}
\label{sec:ls}
To search for periodicity in the radio data, we first created a light curve in both Stokes\,V and Stokes\,I. We binned the data used for the dynamic spectra in 1-hour time bins and averaged over the entire bandwidth. Such averaging often underestimates the maximum flux density, since some bursts are narrow in frequency, but it is the most impartial approach to forming the light curves considering the variety of emission we have detected. The Stokes\,V light curve is more reliable than the Stokes\,I light curve since the Stokes\,V data are not affected by sidelobe noise.

\begin{figure*}
    \centering
    \begin{subfigure}[t]{0.49\textwidth}
    \captionsetup{width=.9\textwidth}
    \centering
    \includegraphics[width=\columnwidth]{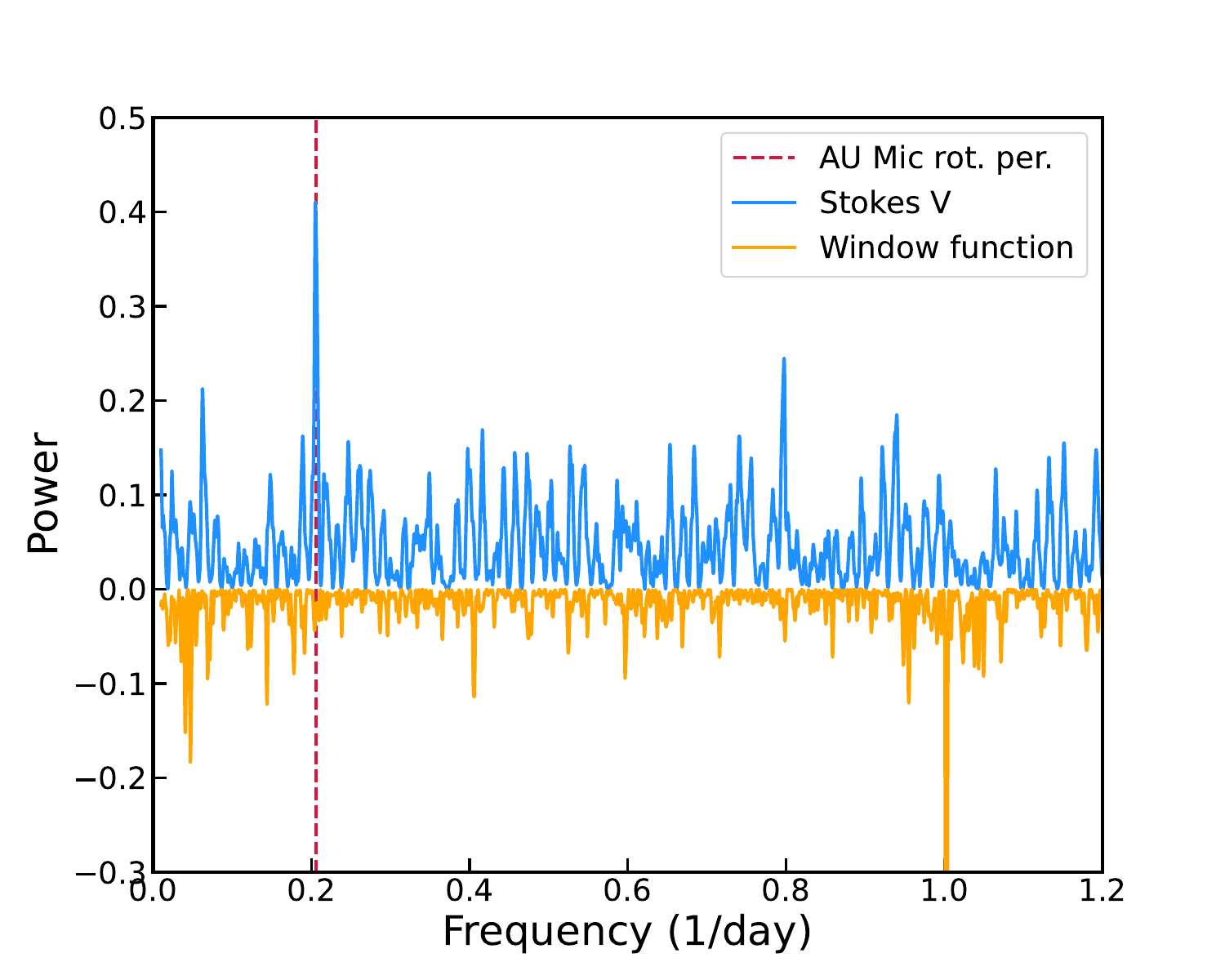}
    \caption{}
    \label{fig:LS_V}
\end{subfigure}
    \begin{subfigure}[t]{0.49\textwidth}
    \captionsetup{width=.9\textwidth}
    \centering
    \includegraphics[width=\columnwidth]{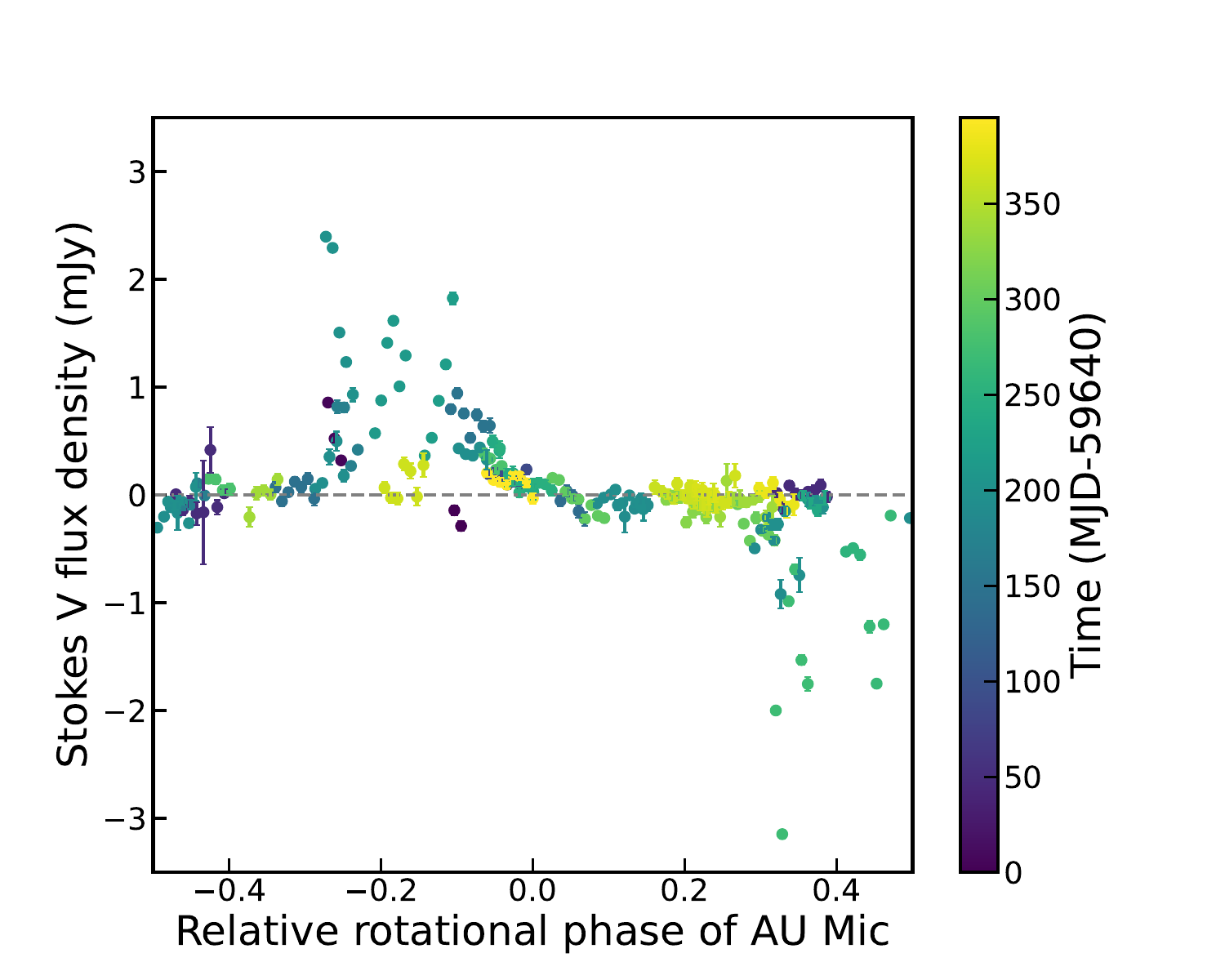}
    \caption{}
    \label{fig:wrapped_lightcurve}
\end{subfigure}
\caption{Results of the periodicity analysis on the Stokes\,V light curve. Panel a: Lomb-Scargle periodogram of our Stokes\,V light curve, as shown in blue. The red line marks the rotation period of AU\,Mic \citep{aumic_b}. There are no strong peaks associated with the periods of the planets. The orange line communicates the Lomb-Scargle periodogram of the window function, scaled to be readable. Panel b: Stokes\,V light curve including all radio data, phase folded to the rotation period of AU\,Mic of 4.86 days. The data points represent the flux density averaged over the entire bandwidth in 1-hour time bins. The colour scale represents the time of observing. The zero point is arbitrary, set to the last observation in our campaign at MJD 60035.1.}
\end{figure*}

The first method we use to search for periodicity in the light curve is the Lomb-Scargle periodogram as implemented in \textsc{Astropy} \citep{lomb, scargle, 2013A&A...558A..33A}. Fig.\,\ref{fig:LS_V} shows the Lomb-Scargle periodogram of the entire Stokes\,V light curve, with a vertical dashed line indicating the frequency of the rotation period of AU\,Mic. We observe a strong peak at a period of 4.876 days, which is conspicuously close to the stellar rotation period of 4.86$\pm$0.01 days, as reported from TESS measurements \citep{aumic_b}. This peak does not exist in the window function. The false alarm probability (FAP) of such a peak between a frequency of 0.01 and 2.0 per day, calculated by randomly shuffling the observations, is 0.0019\%. The other peaks in the periodogram do not correspond to any known physical periods of interest in the system. Notably, we do not identify any peaks corresponding to the orbital periods of the planets (8.46 and 18.9\,days, \citet{aumic_b} and \citet{aumic_c}, or their synodic periods relative to the stellar rotation (11.4 and 6.54\,days).

\begin{figure*}
    \centering
    \begin{subfigure}[t]{0.49\textwidth}
    \captionsetup{width=.9\textwidth}
    \centering
    \includegraphics[width=\columnwidth]{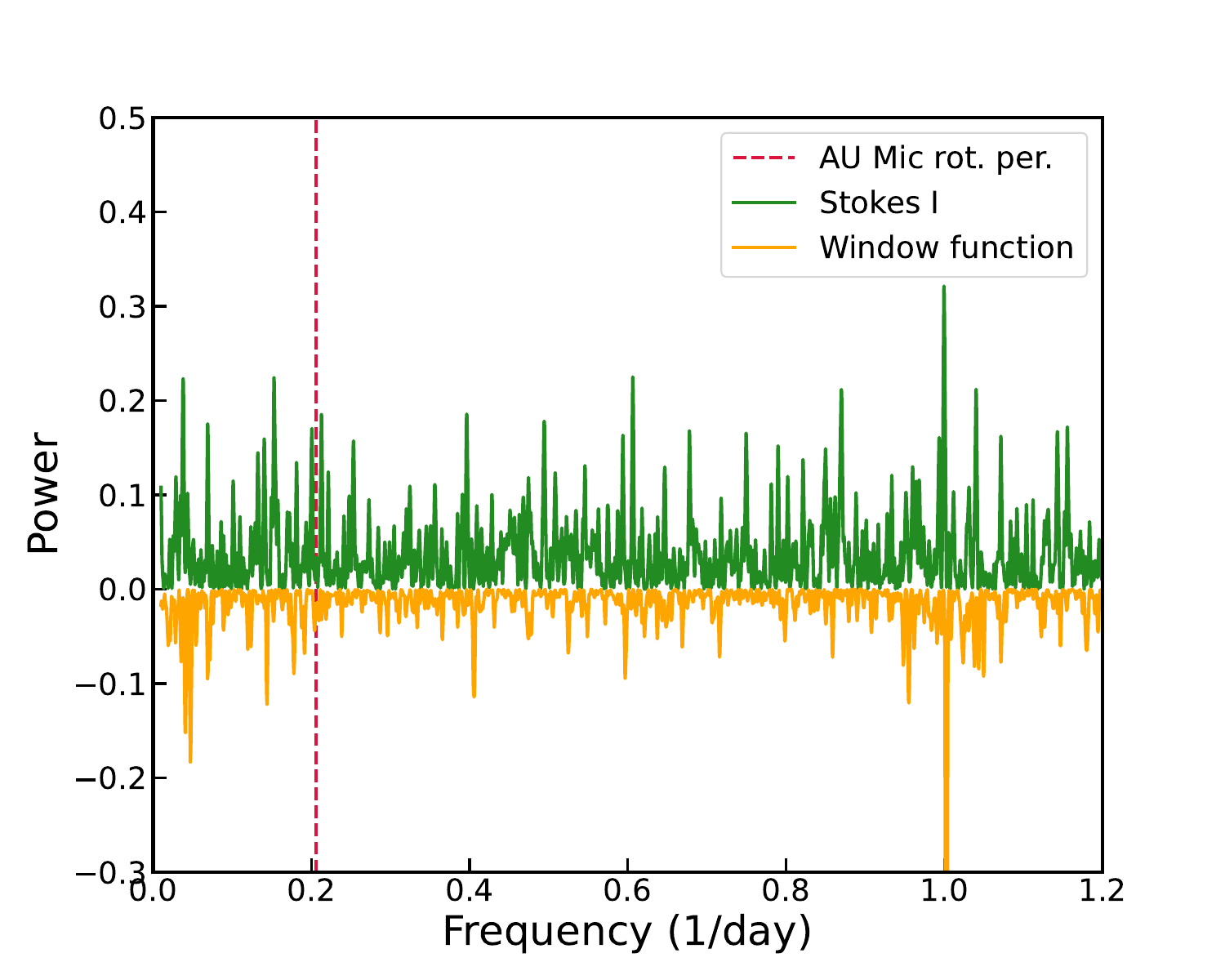}
    \caption{}
    \label{fig:LS_I}
\end{subfigure}
    \begin{subfigure}[t]{0.49\textwidth}
    \captionsetup{width=.9\textwidth}
    \centering
    \includegraphics[width=\columnwidth]{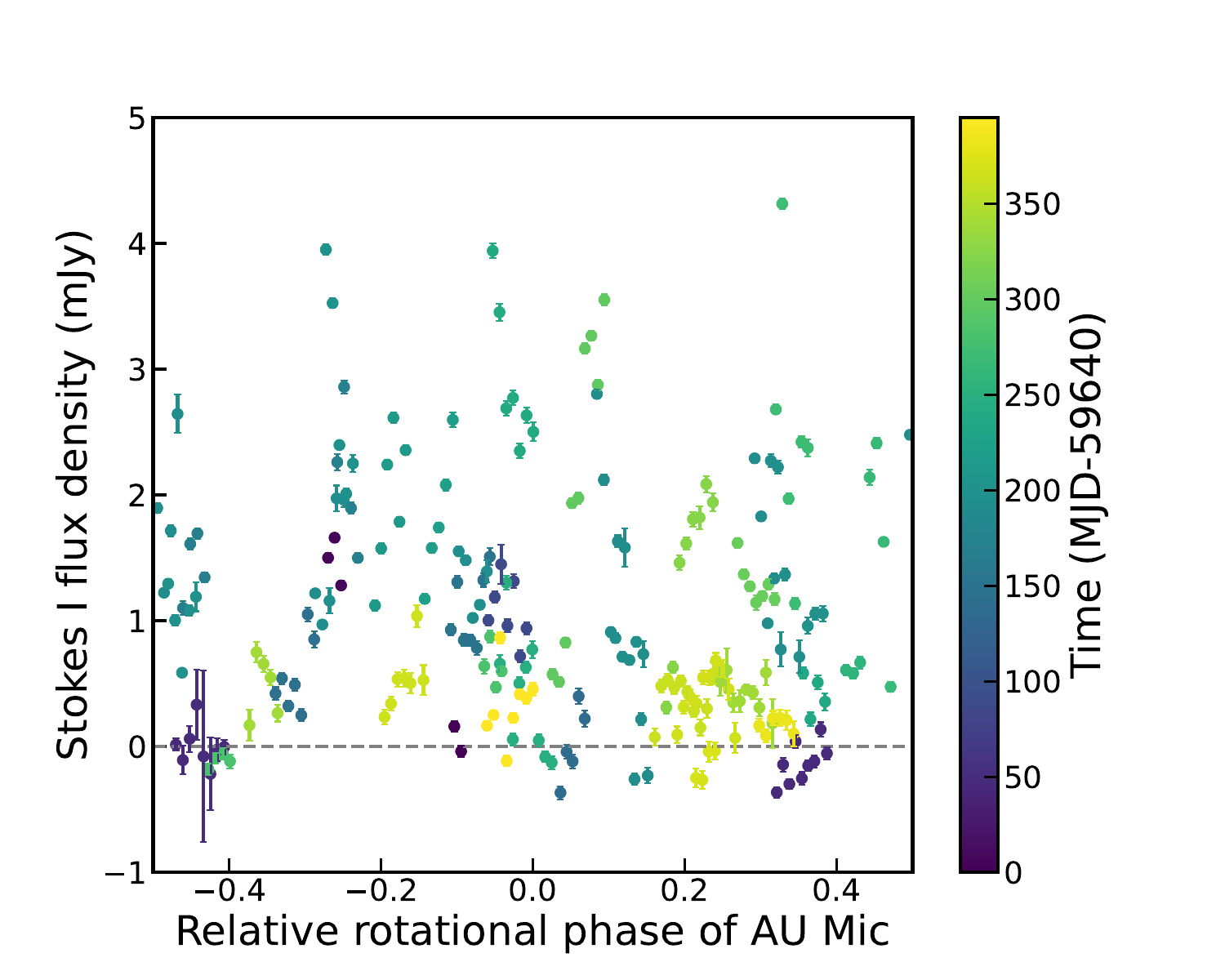}
    \caption{}
    \label{fig:wrapped_lightcurve_I}
\end{subfigure}
\caption{Results of the periodicity analysis on the Stokes\,I light curve. Panel a: Lomb-Scargle periodogram of our Stokes\,I light curve. The red line marks the rotation period of AU\,Mic \citep{aumic_b}. There are no strong peaks associated with the periods of the planets. The orange line shows the Lomb-Scargle periodogram of the window function, scaled to fit the plot. Panel b: Stokes\,I light curve including all radio data, phase wrapped to the rotation period of AU\,Mic of 4.86 days. The data points represent the flux density averaged over the entire bandwidth in 1-hour time bins. The colour scale represents the time of observing. The zero point is arbitrary, set to the last observation in our campaign at MJD 60035.1.}
\end{figure*}

\begin{figure}
    \centering
    \includegraphics[width=\columnwidth]{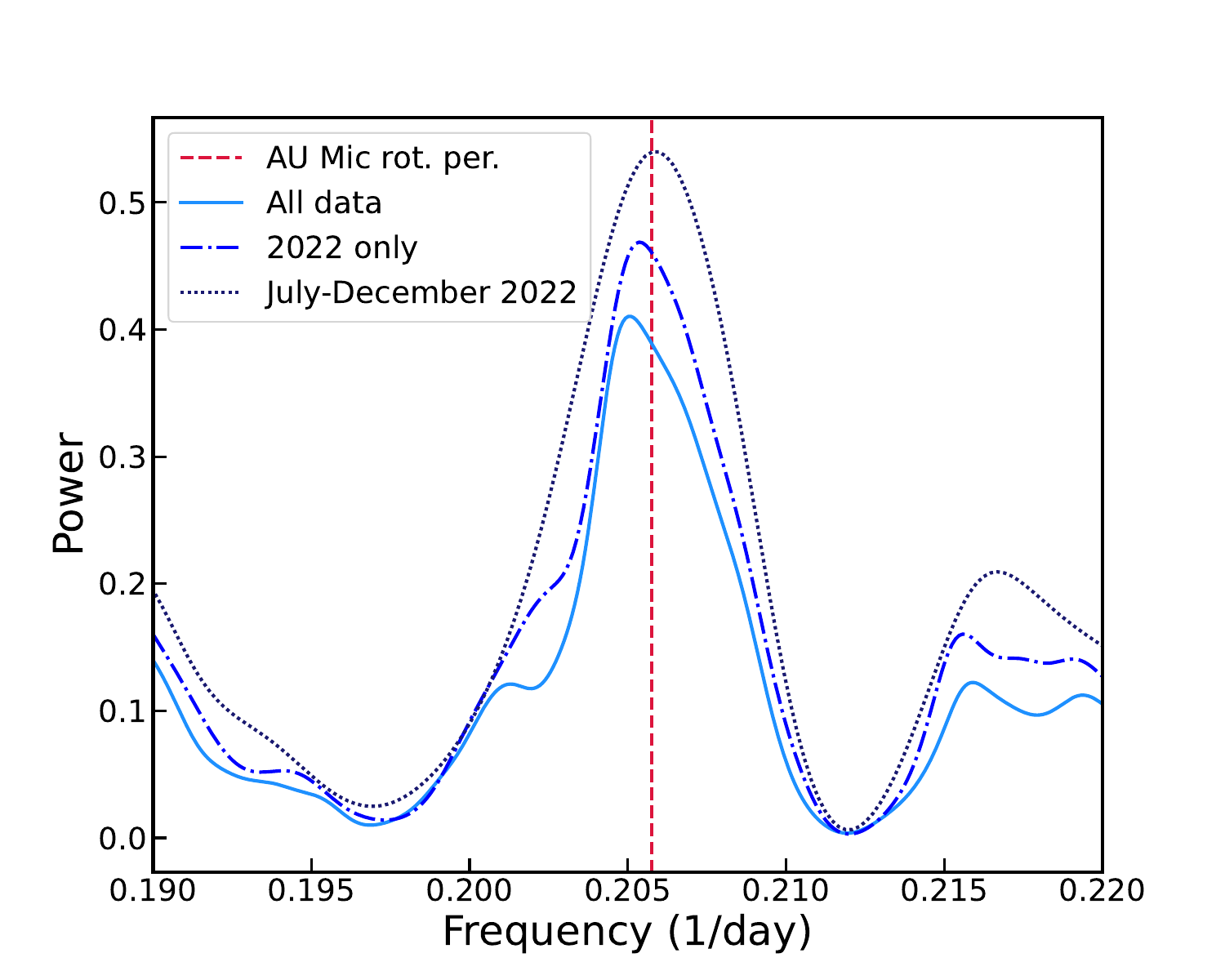}
    \caption{Zoom-in of the Lomb-Scargle periodogram of three different selections of data. The light blue solid line includes all Stokes\,V data. The blue dash-dotted line includes only data takes in 2022, and the dark blue dotted line includes data from the period between 2022-07-30 and 2022-12-31, during which AU\,Mic was most active in our observations.}
    \label{fig:LS_zoom}
\end{figure}

Additionally, we split the data in several different ways to test if the detected periodicity is always present. We are essentially performing jackknife tests to assess the reliability of the periodicity detection and, potentially, provide insight into what radio emission mechanisms are operating. For the first cut, we use only the data from 2022, excluding the data sets from 2023. With this cut, the periodogram looks almost identical to that of the full data set, with a very similar peak to the original. The peak is shown in Fig.\,\ref{fig:LS_zoom}. The period associated with this peak is 4.869\,days. The second cut we do is using only the data from 2022-07-31 to 2022-12-31. This cut is motivated by the fact we detect the majority of bursts in this time baseline, compared to the much smaller number of bursts in the rest of the data. With this selection, the peak in the periodogram is at 4.860\,days, consistent with the measured rotation period of 4.86 days, within our precision. The three periods found with the different cuts are all very similar and the peaks in the periodograms show complete overlap. We therefore conclude that the detection of periodicity is robust and present over the entire monitoring period.

We also phase-wrap the Stokes\,V light curve to 4.86\,days, shown in Fig.\,\ref{fig:wrapped_lightcurve}. The periodicity in the emission is clearly visible by the two peaks in emission, offset approximately 180 degrees in phase.

With a clear periodicity in Stokes\,V data, we repeated the same analysis on the Stokes\,I light curve. However, as shown in Fig.\,\ref{fig:LS_I}, there is no peak at the rotation period of AU\,Mic. In fact, the only significant peak in the periodogram is at 1~sidereal day, which is an artefact of the window function. Although the Stokes\,I data are slightly less reliable due to contamination from nearby sources, the periodic signal completely disappearing implies that the added noise is not the only cause.
Indeed, when we wrap the Stokes\,I light curve to the rotation period of AU\,Mic, shown in Fig.\,\ref{fig:wrapped_lightcurve_I}, it is clear no phase is favoured for detecting the source.

\begin{figure*}
    \centering
    \begin{subfigure}[t]{0.49\textwidth}
    \captionsetup{width=.9\textwidth}
    \centering
    \includegraphics[width=\columnwidth]{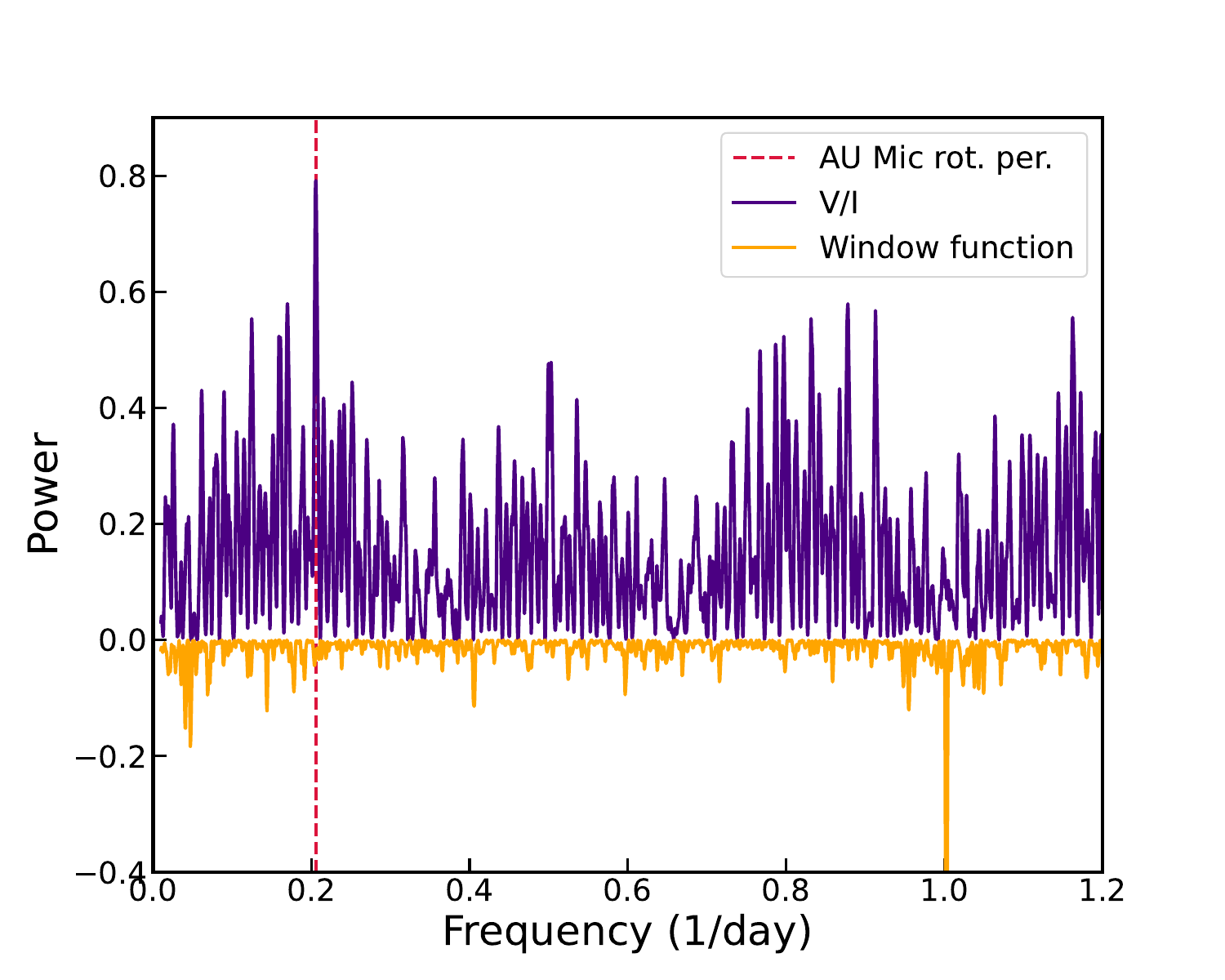}
    \caption{}
    \label{fig:LS_VI}
\end{subfigure}
    \begin{subfigure}[t]{0.49\textwidth}
    \captionsetup{width=.9\textwidth}
    \centering
    \includegraphics[width=\columnwidth]{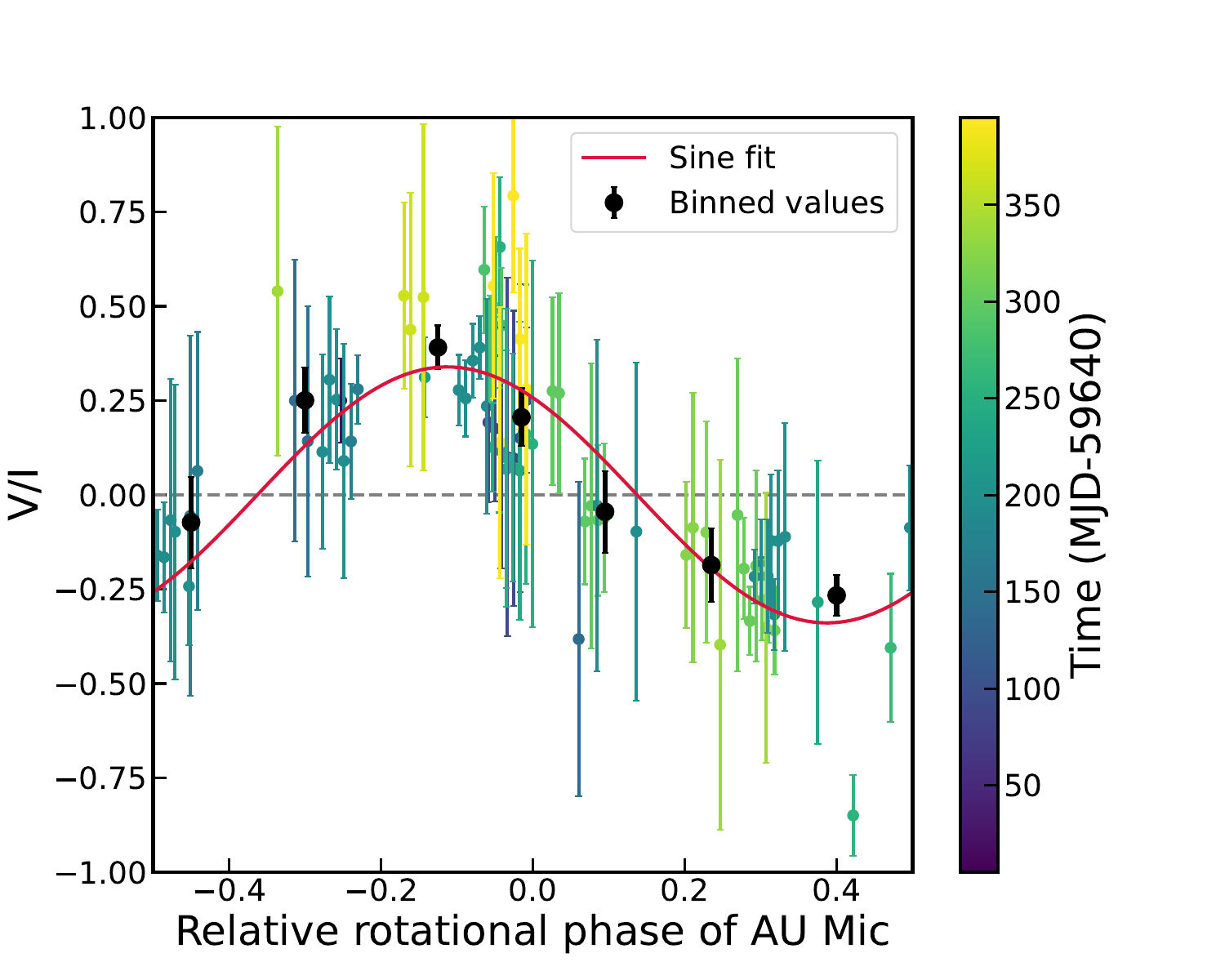}
    \caption{}
    \label{fig:lc_vi}
    
\end{subfigure}
\caption{Results of the periodicity analysis on the polarisation fraction over time, selecting only data points that are physical, have an error bar smaller than 0.5, and have a flux density below 0.5\,mJy to remove the bursts. Panel a: Lomb-Scargle periodogram of the polarisation fraction. The red line marks the rotation period of AU\,Mic \citep{aumic_b}. The orange line shows the Lomb-Scargle periodogram of the window function, scaled to fit the plot. Panel b: Circular polarisation fraction of emission as a function of phase. The colour scale represents the time of observing. The black data points show the binned data, where the bins are chosen such that each bin contains at least 7 data points. The red line shows a sine fit to the phase-folded data with a fixed period of 4.86\,days, as discussed in Sec.\,\ref{sec:broadband}. The zero point is arbitrary, set to the last observation in our campaign at MJD 60035.1.}
\label{fig:vi}
\end{figure*}

Even in epochs without bursts, we often detect faint Stokes\,V emission. This can be seen in Fig.\,\ref{fig:wrapped_lightcurve}, for example around a phase of zero. To determine the nature of this non-burst emission, we repeat the analysis above for the polarisation fraction, excluding all points with an absolute Stokes\,V flux of more than 0.5\,mJy, to remove most of the burst emission. We also exclude all points with an absolute polarisation fraction above one, as these points are unphysical and are a product of low signal-to-noise measurements. The Lomb-Scargle periodogram of this light curve is shown in Fig.\,\ref{fig:LS_VI}. We detect a periodicity in the polarisation fraction at the rotation period of AU\,Mic, albeit at a lower significance than the periodicity detected in the Stokes\,V data. The light curve wrapped to the rotation period of AU\,Mic is shown in Fig.\,\ref{fig:lc_vi}. We also plot the data binned into 7 bins, where the size of the bins is chosen in such a way that each bin contains at least 7 data points.

To ascertain the presence of a periodicity in the polarised fraction, we fitted both a straight line and a sine to the polarisation fraction over time. We fitted the models to the data using MultiNest \citep{Feroz2013, 2014A&A...564A.125B}, using the logarithm of the Bayesian evidence to determine which model is a better fit. Fitting a straight line to the data results in an almost perfectly horizontal line, with a slope of (1.5$\pm$11)$\times10^{-6}$ and an offset at zero of $0.1\pm0.7$. The associated log evidence value is $-119\pm0.2$.

We also fitted a simple sine to the data, given by 
\begin{equation}
\frac{\mathrm{V}}{\mathrm{I}}=a\sin(\frac{2 \pi t}{P} +t_0),
\end{equation}
where $a$ is the amplitude, $t$ is the time of observing, $P$ is the period and $t_0$ is the zero point.
We find a period of 4.863$\pm$0.004\,days, in good agreement with both the Lomb-Scargle periodogram of this dataset and the previously measured rotation period of AU\,Mic \citep{aumic_b}. The best-fit amplitude is 0.3$\pm$0.35 and $t_0$ is 3.2$\pm$2.1. The error bars on the last two parameters are large, possibly caused by the long intervals between the observations. The log evidence value for this model is 4.5$\pm$0.17. The difference between the two models is therefore 123.5. Using an updated version of the Jefferys scale \citep[e.g.][]{Kass1995, Scaife2012}, a difference of 3 in log evidence values is enough to determine that one model is better than the other. We can therefore conclude that the sinusoidal model is a significantly better fit than the model that suggests no change in polarisation fraction with phase -- implying periodicity.

\subsection{Interpretation of the periodicity in the burst emission} 
When the Stokes\,V light curve is phase wrapped to the stellar rotation period, as shown in Fig.\,\ref{fig:wrapped_lightcurve}, the positive bursts always occur at the same phase, shifted roughly 180 degrees in phase compared to the negative bursts. This is consistent with emission with a beaming effect, produced at the northern and southern poles. The only radio emission mechanism that is known to produce such polarisation and periodicity is the ECMI mechanism.

We find that the periodicity in our radio data is consistent with the rotation rate measured with TESS, which is most likely measured close to the equator \citep{aumic_b}. We interpret this to mean that the large-scale magnetic field that drives the ECMI bursts is rotating at a rotation velocity very close to that of the equator. The large-scale magnetic field is generated in a dynamo region in the interior of the star. Therefore, the rotation period we measure with the radio data likely represents the rotation velocity of the interior dynamo region.

If the bursts are due to ECMI in a large-scale dipolar field, we can determine the stellar magnetic field geometry. AU\,Mic's rotational axis is inclined at $\approx 90^\circ$ relative to the line of sight, which further simplifies our interpretation. We adopt an auroral ring model wherein ECMI emission originates in northern and southern auroral rings at a certain magnetic latitude, although due to beaming, not all field lines on the rings will be visible to an observer at a given time. Generally speaking, if the magnetic obliquity is sufficiently large, we expect emission from the northern auroral ring to intersect with our line of sight twice per rotation. The same applies to the southern auroral ring, yielding two additional radio pulses with the opposite handedness compared to the pulses from the northern auroral ring. The relative position of the four emission visibility zones along the rotation phase axis depends on the magnetic obliquity and the opening angle of the cone of emission. 

In the emission pattern of AU\,Mic we see that the emission largely appears in just two visibility zones of opposite handedness, with each zone confined to a region in phase of at most 0.25. 
This suggests that the two anticipated visibility zones per magnetic hemisphere must overlap or be close to one another in rotation phase. This can happen when the opening angle of the ECMI cone, magnetic obliquity and the magnetic co-latitude of the auroral rings together add approximately to the inclination of $\approx 90^\circ$. In this geometry, the radio bursts become visible when either magnetic pole of the star is closely aligned with the meridional plane (i.e. the plane formed by the line of sight vector and the stellar rotation axis).

\begin{figure}
    \centering
    \includegraphics[width=\columnwidth]{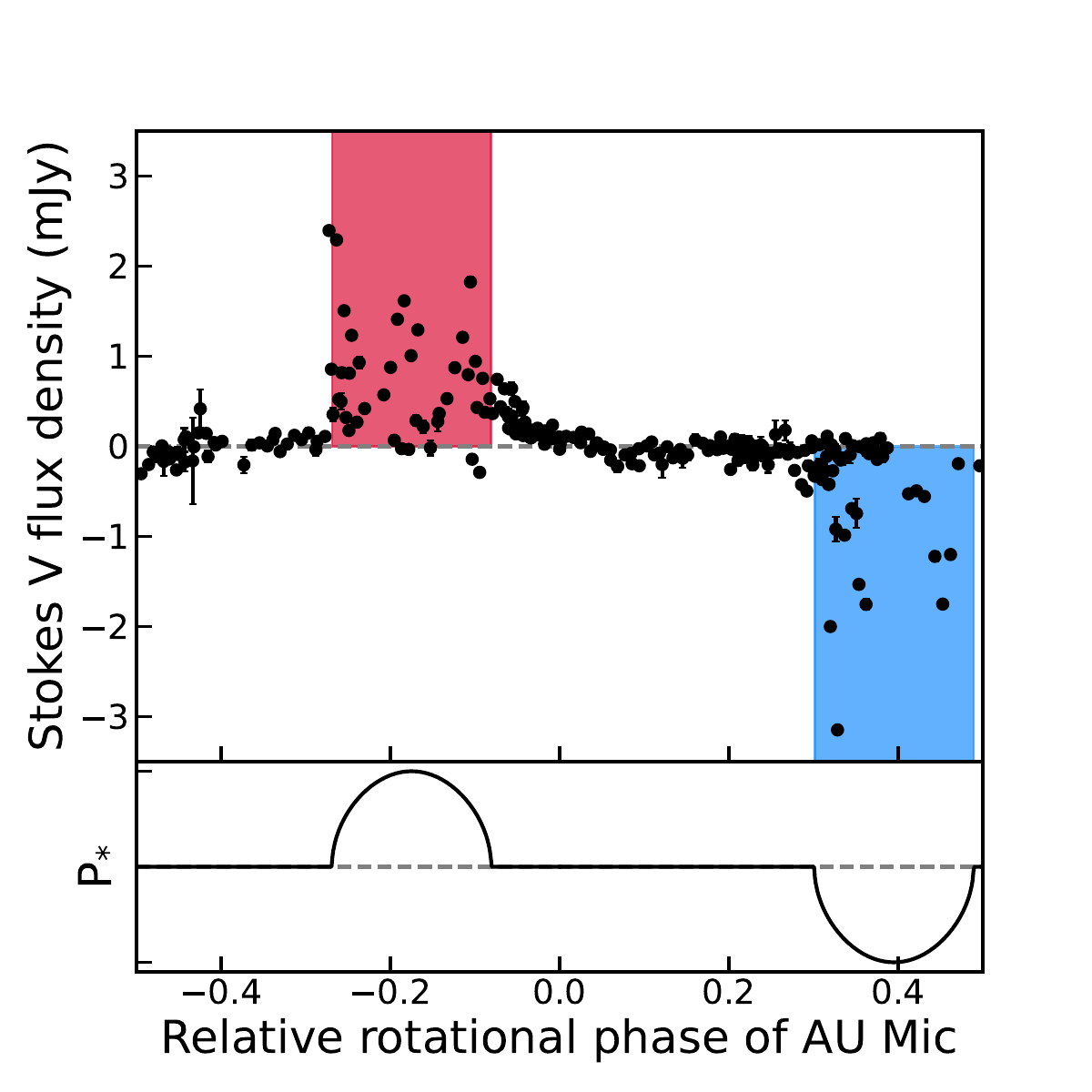}
    \caption{Comparison of the Stokes\,V light curve to an auroral ring model. The top plot shows the Stokes\,V light curve including all radio data, phase folded to the rotation period of AU\,Mic of 4.86 days. The data points represent the flux density averaged over the entire bandwidth in 1-hour time bins. The two shaded regions show the two periods of burst visibility, where the red region shows when positive bursts can be produced and the blue region shows the same for negative bursts. The bottom plot shows a model of the visibility of the auroral rings, where P$_{*}$ represents the visibility of the auroral ring, akin to the probability of detecting a burst, multiplied by the predicted sign of emission.}
    \label{fig:aur}
\end{figure}

To test this geometric interpretation, we numerically computed the visibility of ECMI emission from an auroral ring model, discussed in detail in Appendix\,\ref{app:aur}. We note that this model does not represent a prediction of the flux density of a burst at that moment, but rather whether it is possible for emission to be visible at any given time. The emission is not exactly 180$\degree$ apart in phase, as is expected for a perfect dipole. From ZDI maps \citep[e.g.][]{zdi_aumic}, we know that the magnetic field of AU\,Mic is not a perfect dipole, but instead has some phase offset between the two poles. We therefore model the two periods of bursts separately, with a phase offset between the two poles. Both are modelled with a cone opening angle of 65$\degree$, a cone thickness of 5$\degree$, an inclination of 90$\degree$, a magnetic obliquity of 20$\degree$, and an auroral ring magnetic latitude of 4$\degree$. The phase difference between the positive and the negative peak is 0.57. We show this model in Fig.\,\ref{fig:aur}, where the top plot shows the phase-wrapped Stokes\,V light curve with the phases at which bursts can be visible shaded. The bottom plot shows the full model as explained in Appendix\,\ref{app:aur}.

{Naturally, it is worth validating the inferred geometry of the auroral ring in the context of the magnetic field strength of AU Mic and the observing frequency. The model described in Appendix~\ref{app:aur} does not depend on either of these parameters. However, one can utilise Eq.~\ref{eq:dipole field strength} in combination with the inferred magnetic co-latitude of $4\degr$ for auroral ring to determine the minimum field strength for the dipole which places the ring above the stellar surface.}

For the ring to sit above the stellar surface, the field strength at the magnetic poles must be such that the cyclotron frequency at the ring location equals the observing frequency $\nu$. Enforcing this along with a the requirement that the radial distance of the ring from the centre of the star exceeds the stellar radius, one can show from Eq.~\ref{eq:dipole field strength} that the field strength of the dipole must be:
\begin{equation}
B_\star > \frac{\nu}{1.4~\mathrm{MHz} \times (1 + 3\cos^2\theta)^{1/2}}~\mathrm{G} .
\end{equation}
Plugging in for the observing frequency with $\nu = 2$~GHz and a magnetic co-latitude of $\theta = 4\degr$, we find that the field strength of the dipole must be greater than 717~G for the auroral ring to sit above the surface. This is in good agreement with the results from \citet{zdi_aumic}, who estimate a dipole field strength of 660~G for AU Mic in 2022.

We note that there is no periodicity with the orbital or synodic periods of the planets. At first glance, this implies that the planets are not inducing SPI emission on the planet. However, in the geometry determined above, a planet would only be able to induce radio bursts when it is near conjunction and the magnetic pole of the star is pointing towards Earth. Although not all bursts can be caused in this way, we cannot rule out that the planets may be inducing some of the bursts.
Alternatively, we are not seeing any SPI-induced bursts because the planets are outside the Alfv\'{e}n surface of AU\,Mic. This suggestion will be explored further in future work.

\subsection{Interpretation of the broadband emission}
\label{sec:broadband}
We detect bursts in 16 of 42 epochs. The epochs where we detect no bursts generally do show Type\,A emission. In Sec.\,\ref{sec:ls}, we searched the data from 2023, which contains only Type\,A emission and no bursts, for periodicity and still found the same signal.
This implies that the Stokes\,V component of the Type\,A emission is periodic. When performing an identical analysis on the Stokes\,I data, the periodicity does not appear, nor does it look periodic in a light curve. The polarisation fraction is periodic with the rotation rate, as shown in Fig.\,\ref{fig:vi}. The handedness of the polarised emission is generally the same for the bursts and the Type\,A emission. 
Taken all together, this implies that the emission is produced on the same hemisphere as the ECMI emission producing the bursts, but through a different emission mechanism. As described in Section\,\ref{sec:em}, the emission mechanism is likely to be {gyromagnetic} emission, based on the polarisation fraction and the time-frequency structure. In this section, we present a qualitative analysis of the evolution of {gyromagnetic} emission with the rotation phase of the star.

The handedness of the emission is determined by the direction of the magnetic field in the region where it is emitted. On the Sun, the emission is produced in small loops on the surface, where the direction of the magnetic field is determined by the local structure. AU\,Mic, in comparison, has a much stronger magnetic field that is less variable in time. The magnetic dipole strength of AU\,Mic is at least 1000 times stronger than that of the Sun. We conjecture that the electrons may be energised in small loops near the surface, but spend a substantial fraction of their time trapped within the large-scale field Therefore, their emission could bear the signature of the large-scale field. With this configuration, the mode of emission is determined by the hemisphere on which the electrons are located.

Assuming the sense of polarisation of the {gyromagnetic} emission is therefore mostly determined by the magnetic hemisphere of emission, the polarisation fraction of the emission can be periodic as long as the magnetic obliquity is not zero. With a magnetic obliquity larger than zero, the projected area of each hemisphere changes as a function of phase. Since we cannot resolve the surface of the star, the polarisation fraction we observe is a result of averaging over the entire stellar disk. If one pole is pointed towards us, the corresponding hemisphere will have a larger projected surface than the other. Assuming both hemispheres emit equally, this would result in a net circular polarisation of the emission, which will vary with phase.
If there is no magnetic obliquity, the projected area of the northern and southern magnetic hemisphere does not change, and the polarisation fraction should be constant.

We have already fit a sinusoid to the data to determine the periodicity and amplitude of the variation in polarisation fraction earlier in this section. However, the error bars on these values are very large. Since we have confirmed the polarisation fraction is periodic, we now fit a sinusoid to the phase-folded data using \texttt{emcee} \citep{ForemanMackey2013} to more accurately determine the amplitude while keeping the period fixed at 4.86\,days. We find an amplitude of 0.34$\pm$0.034. To explain this variation with the uniform model explained above, we require that the difference in the projected area of the northern and southern magnetic hemispheres must be the same as the circular polarisation fraction if the emission is 100\% circularly polarised at its origin. The magnetic obliquity required to create this projected area difference is 20$\pm$2\degree. If the emission is originally not 100\% circularly polarised, the magnetic obliquity has to be larger than the value given above.

Comparing the magnetic obliquity we find to that found in ZDI maps \citep{zdi_aumic}, the two different methods to determine the obliquity agree. However, we note that our value has been determined assuming completely uniform conditions across the entire stellar surface, except for the change in direction of the magnetic field. Using different structures for the structure will most likely lead to different values. We intend to refine our model to provide more stringent constraints in future work.

To create the required polarisation fractions of each hemisphere to average out to the observed circular polarisation fraction of at most 34\%, the emission from each hemisphere must have a high circular polarisation fraction. The {gyromagnetic} emission must therefore be produced at a low harmonic number and/or produced by a highly anisotropic electron population. In either case, the brightness temperature of the emission cannot exceed the kinetic temperature of emitting electrons. Even mildly relativistic electrons have a kinetic temperature of $\sim10^9$\,K. The brightness temperature of the broadband emission we observe is also of the order of $10^9$\,K. Therefore, we conclude that {gyromagnetic} emission is likely producing the broadband emission we observe, as well as the periodicity in the polarisation fraction. 

\subsection{Determining the mode of emission}
If we can confirm which magnetic hemisphere the emission originates on, we can determine whether the emission is in the magnetoionic o- or x-mode. We note that, in the IAU convention, if the magnetic field vector points towards the observer, o-mode emission is defined as negative Stokes\,V flux, and x-mode emission as positive Stokes\,V flux. 

We can identify the magnetic hemisphere of emission by comparing the rotation phase at which AU\,Mic's radio bursts arrive with the orientation of the magnetic axis at the time of observation as predicted by the ZDI maps of \citet{zdi_aumic}. The zero-point of the ZDI maps presented in \citet{zdi_aumic} is set at JD\,2459000. The difference in phase between this zero-point and the zero-point we use in this work is 0.09. When using the zero point used for the ZDI maps to phase-wrap our radio data, we see that the right-handed bursts occur almost exactly at the phase where the magnetic north pole from the 2022 ZDI map is in the meridional plane, as predicted by our simple auroral-ring model. The magnetic south pole on this map is offset by approximately 55\degree when compared to the left-handed radio emission. However, when we compare the radio data to the ZDI map from 2021 (which is one of the largest datasets), we find that the positive bursts still align with the magnetic north pole, while the negative bursts now also align with the magnetic south pole. Taken together, we conclude that the right-handed emission most likely originates on the magnetic north pole, and the left-handed emission on the south pole. In this case, the emission is in the magneto-ionic x-mode.

{Gyromagnetic} emission in the optically thick regime is expected to be polarised in the sense of the o-mode, and polarised in the sense of the x-mode in the optically thin regime. Since the spectral index of the broadband emission is generally positive, but not steep, the emission is most likely in the transition region between optically thin and optically thick emission. The mode of emission could therefore be either o- or x-mode. We expect the {gyromagnetic emission's} polarisation to be dominated by the magnetic hemisphere that is also producing bursts at a given phase. Since the handedness of the bursts and the broadband emission is generally consistent, we expect both emission mechanisms to be in the same magneto-ionic mode.

\subsection{Exceptions}
The periodicity in emission described in this section can be seen in both the bursts and the broadband emission. However, not all emission fits the periodicity.

All Type\,B, C, D, and F bursts fit with the periodic modulation of emission. The handedness of emission is fully determined by the phase at which it is observed. On the other hand, the Type\,E shot bursts do not seem to follow this pattern. We have only 2 detections of this type of emission at these frequencies, both with right-handed circular polarisation. For the first Type\,E burst, this is expected considering the phase of rotation. However, for the second burst, its phase suggests the emission should be left-handed circularly polarised. The broadband emission in this observation is left-handed, but the burst is right-handed. The reason for this could be that this particular burst did not originate on the northern magnetic hemisphere with the other bursts at that phase, but instead on the southern hemisphere in a local event.

In general, the Type\,A emission aligns with the expected handedness based on the phase. There is only one exception, which is shown in Fig\,\ref{fig:unconfinec}. The Type\,A emission in this epoch is left-handed, when the periodicity predicts it should be right-handed. A possible explanation for the change in handedness is that the southern hemisphere was unusually active during this observation, increasing the brightness of the emission from the southern hemisphere relative to the northern hemisphere, therefore changing the averaged polarisation fraction.

\section{Changes in radio activity}
\label{sec:variation}
Targeting AU\,Mic regularly for a period of over a year allows us to not only probe the short-term patterns presented in the previous section, but any possible long-term variations as well. In this section, we present an analysis of a potential long-term variation in our data set.

In 2022, we observed AU\,Mic 33 times, with a burst detected in 16 out of 33 epochs. Based on the phase-wrapped light curve, we expect the star to be on for around 45-50\% of the time, which agrees well with the observed fraction in 2022. However, the observations in 2023 do not follow the same trend. In 9 observations, we do not detect a single burst. Considering the rotational phase of these observations, most were unfortunately scheduled during the "off" phases. However, two observations were scheduled within the "on" phase, but no burst was detected. Notably, out of the nine observations in 2023, only one shows any detectable Stokes\,V emission at all. Comparing this to the baseline trend of 2022, the number of non-detections is remarkable.

Assuming the burst emission is produced through ECMI, as suggested by the periodicity, there are a few reasons for the radio bursts to stop. The first option is that the magnetic pole is shifting compared to the rotation axis, changing the direction of the beamed emission. Magnetic fields of M\,dwarfs changing on a timescale of a year is not unheard of, as shown by \citet{2023A&A...676A..56B}.
From the ZDI maps of AU\,Mic over time, it does look like the magnetic pole is not completely static. It appears to be precessing, but such motion should not affect the amount of time our line of sight intersects with the emission cones significantly. The precession would only change the phase at which we see the bursts. Based on the complete lack of any bursts in the data taken in 2023, a change in the magnetic field is unlikely to be the cause of the sudden stop in bursts.

A second option is that the strength of the dipole is changing. In our observations, we see bursts across the band, from 1 to 3\,GHz. This suggests a local magnetic field strength of 0.5-1 kG. Changing the field strength to lower values would move the cyclotron frequency in those regions out of our band.
However, we do not see a significant decrease or increase in the burst frequency in the period leading up to the inactive period. The frequency at which the bursts occur seems to vary at random, with no tendency towards lower or higher values as time goes on.

Since the ECMI mechanism still requires energetic electrons to produce radio emission, decreasing the supply of high-energy electrons would dim the radio bursts as well as type-A emission. Although the source of high-energy electrons is not known, a possibility is that they are provided by the flares on the star. If this is the case, a decrease in the frequency of energetic flares might be responsible for the inactive period we see. This would also agree with the decrease in the brightness of broadband emission at the end of our observing campaign, which can be seen in Fig.\,\ref{fig:wrapped_lightcurve_I}. Unfortunately, we do not have access to enough contemporaneous photometric data in this period to test this theory. 

Finally, it is possible that AU\,Mic is still producing radio bursts, but we have not observed at the right time in the last three months of our observing campaign. Although it seems unlikely with the periodicity seen in the first 10 months, it cannot be ruled out considering our sample size.

\section{Conclusions}
\label{sec:concl}
In this work, we have presented the results of our unrivalled radio monitoring campaign on AU\,Mic. With over 250\,h of observations, we have catalogued and described six different types of emission produced on the star. These can be split into two overarching categories: broadband emission and radio bursts. We develop a classification scheme for the phenomenology of the radio emission observed on AU\,Mic that hopefully can be useful for future studies of radio stars. 

The detected circularly polarised emission is periodic, with a best-fit period close to the rotation period of the star. The total intensity of emission is found not to be periodic. There is no detection of periodicity at the orbital periods of the planets in the system or their synodic periods, implying we find no evidence for radio emission from a star-planet interaction in the AU\,Mic system. This could be because the planets are not within the Alfv\'{e}n surface of AU\,Mic, which will be explored in detail in future work.

The radio bursts we detect on AU\,Mic are strongly circularly polarised, implying that they must be caused by a coherent emission mechanism. The fact that the bursts are periodic with the rotation of the star confirms that this emission is produced through the ECMI mechanism, close to the magnetic poles. We theorise that the emission is likely generated in an auroral ring, similar to the emission seen on Jupiter, and present a qualitative model to determine the burst probability at a given rotational phase.

The broadband component of emission is likely not caused by ECMI, but instead by low-harmonic {gyromagnetic} emission. The total intensity of this emission is not periodic, but the circular polarisation fraction does vary with the rotation phase of the star. We present a qualitative model to describe this emission pattern, with {gyromagnetic} emission being produced across the entire surface of the star and the polarisation fraction being determined by averaging over the entire stellar disk.
Although our model is simple, we can estimate that the magnetic obliquity required to produce this polarisation fraction must be larger than 20\degree. We expect that further study of the emission on AU\,Mic at higher frequencies around 10\,GHz, combined with more sophisticated modelling, could allow us to confidently determine the structure of the magnetic field.

In the last three months of our campaign, the frequency of detections of circularly polarised emission decreased strongly. Although the number of non-detections is too low to draw significant conclusions, a possible explanation for this is that the constant supply of high-energy electrons by stellar activity has abated.

In conclusion, the radio emission from M\,dwarfs is key to understanding their magnetospheres and environments. Although AU\,Mic is the only star that has been studied at radio frequencies in such detail, other M\,dwarfs have been shown to produce emission that looks qualitatively similar \citep[e.g.][]{villadsen2019}. AU\,Mic is likely not unique. We expect the behaviour we observed should be ubiquitous across M\,dwarfs. To test this prediction, more long-term radio campaigns on different M\,dwarfs are required. In particular, the observations should cover the rotational phase of the star at least two, but ideally three times, to ensure the dataset is robust for detecting periodicity. Campaigns on M\,dwarfs with a variety of masses could also test if there is a transition in the type of radio emission emitted as the interior of the star transitions from having a radiative core to being fully convective. Prime candidates for such campaigns are EQ\,Pegasi and AD\,Leonis, as they are know to produce radio bursts \citep[e.g.][]{2006ApJ...637.1016O,2008ApJ...674.1078O,villadsen2019}, are less massive than AU\,Mic, and have such fast rotation periods that a large amount of telescope time is not needed to fully sample it. Based on our results on AU\,Mic, we predict that these stars likely produce periodic bursts that trace the magnetic field. Such measurements will provide a wealth of information about the dynamo processes generating magnetic fields, while also informing us about the plasma environment around stars.

\section*{Acknowledgements}
SB, HKV and RDK acknowledge funding from the Dutch research council (NWO) under the talent programme (Vidi grant VI.Vidi.203.093). The Australia Telescope Compact Array is part of the Australia Telescope National Facility which is funded by the Australian Government for operation as a National Facility managed by CSIRO. We acknowledge the Gomeroi people as the traditional owners of the Observatory site. BJSP acknowledges and pays respect to the traditional owners of the land on which the University of Queensland is situated, and to their Ancestors and descendants, who continue cultural and spiritual connections to Country. JBC and JCG were supported by projects PID2020-117404GB-C22, funded by MCIN/AEI, CIPROM/2022/64, funded by the Generalitat Valenciana, and by the Astrophysics and High Energy Physics programme by MCIN, with funding from European Union NextGenerationEU (PRTR-C17.I1) and the Generalitat Valenciana through grant ASFAE/2022/018. MPT and LPM acknowledge financial support through grants CEX2021-001131-S and PID2020-117404GB-C21, funded by MCIN/AEI/ 10.13039/501100011033. LPM also acknowledges funding through the grant PRE2020-095421, funded by MCIN/AEI/10.13039/501100011033 and by FSE Investing in your future.

This research made use of NASA's Astrophysics Data System, the \textsc{IPython} package \citep{PER-GRA:2007}; \textsc{SciPy} \citep{scipy}; \textsc{Matplotlib}, a \textsc{Python} library for publication quality graphics \citep{Hunter:2007}; \textsc{Astropy}, a community-developed core \textsc{Python} package for astronomy \citep{2013A&A...558A..33A}; and \textsc{NumPy} \citep{van2011numpy}. 

\subsection*{Data availability}
The data underlying this article are available in the Australia Telescope Online Archive at \url{https://atoa.atnf.csiro.au/} under project IDs C3267, CX513, C3506, and C3469. The code used to create the dynamic spectra in this work is available at \url{https://github.com/SBloot/radio-dynspec}.

\bibliographystyle{aa}
\bibliography{mybib_stars_paper.bib} 

\begin{thebibliography}{80}
\expandafter\ifx\csname natexlab\endcsname\relax\def\natexlab#1{#1}\fi

\bibitem[{{Astropy Collaboration} {et~al.}(2013){Astropy Collaboration}, {Robitaille}, {Tollerud}, {Greenfield}, {Droettboom}, {Bray}, {Aldcroft}, {Davis}, {Ginsburg}, {Price-Whelan}, {Kerzendorf}, {Conley}, {Crighton}, {Barbary}, {Muna}, {Ferguson}, {Grollier}, {Parikh}, {Nair}, {Unther}, {Deil}, {Woillez}, {Conseil}, {Kramer}, {Turner}, {Singer}, {Fox}, {Weaver}, {Zabalza}, {Edwards}, {Azalee Bostroem}, {Burke}, {Casey}, {Crawford}, {Dencheva}, {Ely}, {Jenness}, {Labrie}, {Lim}, {Pierfederici}, {Pontzen}, {Ptak}, {Refsdal}, {Servillat}, \& {Streicher}}]{2013A&A...558A..33A}
{Astropy Collaboration}, {Robitaille}, T.~P., {Tollerud}, E.~J., {et~al.} 2013, \aap, 558, A33

\bibitem[{{Bai} \& {Ramaty}(1979)}]{1979ApJ...227.1072B}
{Bai}, T. \& {Ramaty}, R. 1979, \apj, 227, 1072

\bibitem[{Bastian {et~al.}(2022)Bastian, Cotton, \& Hallinan}]{Bastian_2022}
Bastian, T.~S., Cotton, W.~D., \& Hallinan, G. 2022, \apj, 935, 99

\bibitem[{{Bellotti} {et~al.}(2023){Bellotti}, {Morin}, {Lehmann}, {Folsom}, {Hussain}, {Petit}, {Donati}, {Lavail}, {Carmona}, {Martioli}, {Romano Zaire}, {Alecian}, {Moutou}, {Fouqu{\'e}}, {Alencar}, {Artigau}, {Boisse}, {Bouchy}, {Cadieux}, {Cloutier}, {Cook}, {Delfosse}, {Doyon}, {H{\'e}brard}, {Kochukhov}, \& {Wade}}]{2023A&A...676A..56B}
{Bellotti}, S., {Morin}, J., {Lehmann}, L.~T., {et~al.} 2023, \aap, 676, A56

\bibitem[{{Benz}(2002)}]{2002ASSL..279.....B}
{Benz}, A. 2002, {Plasma Astrophysics, second edition}, Vol. 279

\bibitem[{{Bloomfield} {et~al.}(2002){Bloomfield}, {Mathioudakis}, {Christian}, {Keenan}, \& {Linsky}}]{2002A&A...390..219B}
{Bloomfield}, D.~S., {Mathioudakis}, M., {Christian}, D.~J., {Keenan}, F.~P., \& {Linsky}, J.~L. 2002, \aap, 390, 219

\bibitem[{{Buchner} {et~al.}(2014){Buchner}, {Georgakakis}, {Nandra}, {Hsu}, {Rangel}, {Brightman}, {Merloni}, {Salvato}, {Donley}, \& {Kocevski}}]{2014A&A...564A.125B}
{Buchner}, J., {Georgakakis}, A., {Nandra}, K., {et~al.} 2014, \aap, 564, A125

\bibitem[{{Burgasser} \& {Putman}(2005)}]{2005ApJ...626..486B}
{Burgasser}, A.~J. \& {Putman}, M.~E. 2005, \apj, 626, 486

\bibitem[{{Callingham} {et~al.}(2021{\natexlab{a}}){Callingham}, {Pope}, {Feinstein}, {Vedantham}, {Shimwell}, {Zarka}, {Tasse}, {Lamy}, {Veken}, {Toet}, {Sabater}, {Best}, {van Weeren}, {R{\"o}ttgering}, \& {Ray}}]{crdra}
{Callingham}, J.~R., {Pope}, B.~J.~S., {Feinstein}, A.~D., {et~al.} 2021{\natexlab{a}}, \aap, 648, A13

\bibitem[{{Callingham} {et~al.}(2021{\natexlab{b}}){Callingham}, {Vedantham}, {Shimwell}, {Pope}, {Davis}, {Best}, {Hardcastle}, {R{\"o}ttgering}, {Sabater}, {Tasse}, {van Weeren}, {Williams}, {Zarka}, {de Gasperin}, \& {Drabent}}]{callingham2021}
{Callingham}, J.~R., {Vedantham}, H.~K., {Shimwell}, T.~W., {et~al.} 2021{\natexlab{b}}, Nature Astronomy, 5, 1233

\bibitem[{{CASA Team} {et~al.}(2022){CASA Team}, {Bean}, {Bhatnagar}, {Castro}, {Donovan Meyer}, {Emonts}, {Garcia}, {Garwood}, {Golap}, {Gonzalez Villalba}, {Harris}, {Hayashi}, {Hoskins}, {Hsieh}, {Jagannathan}, {Kawasaki}, {Keimpema}, {Kettenis}, {Lopez}, {Marvil}, {Masters}, {McNichols}, {Mehringer}, {Miel}, {Moellenbrock}, {Montesino}, {Nakazato}, {Ott}, {Petry}, {Pokorny}, {Raba}, {Rau}, {Schiebel}, {Schweighart}, {Sekhar}, {Shimada}, {Small}, {Steeb}, {Sugimoto}, {Suoranta}, {Tsutsumi}, {van Bemmel}, {Verkouter}, {Wells}, {Xiong}, {Szomoru}, {Griffith}, {Glendenning}, \& {Kern}}]{2022PASP..134k4501C}
{CASA Team}, {Bean}, B., {Bhatnagar}, S., {et~al.} 2022, \pasp, 134, 114501

\bibitem[{{Climent} {et~al.}(2020){Climent}, {Guirado}, {Azulay}, {Marcaide}, {Jauncey}, {Lestrade}, \& {Reynolds}}]{2020A&A...641A..90C}
{Climent}, J.~B., {Guirado}, J.~C., {Azulay}, R., {et~al.} 2020, \aap, 641, A90

\bibitem[{{Cox} \& {Gibson}(1985)}]{1985ASSL..116..233C}
{Cox}, J.~J. \& {Gibson}, D.~M. 1985, in Astrophysics and Space Science Library, Vol. 116, Radio Stars, ed. R.~M. {Hjellming} \& D.~M. {Gibson}, 233--236

\bibitem[{{Crosley} \& {Osten}(2018)}]{2018ApJ...862..113C}
{Crosley}, M.~K. \& {Osten}, R.~A. 2018, \apj, 862, 113

\bibitem[{{Donati} {et~al.}(2023){Donati}, {Cristofari}, {Finociety}, {Klein}, {Moutou}, {Gaidos}, {Cadieux}, {Artigau}, {Correia}, {Bou{\'e}}, {Cook}, {Carmona}, {Lehmann}, {Bouvier}, {Martioli}, {Morin}, {Fouqu{\'e}}, {Delfosse}, {Doyon}, {H{\'e}brard}, {Alencar}, {Laskar}, {Arnold}, {Petit}, {K{\'o}sp{\'a}l}, {Vidotto}, {Folsom}, \& {collaboration}}]{zdi_aumic}
{Donati}, J.~F., {Cristofari}, P.~I., {Finociety}, B., {et~al.} 2023, \mnras, 525, 455

\bibitem[{{Dulk}(1985)}]{dulk1985}
{Dulk}, G.~A. 1985, \araa, 23, 169

\bibitem[{{Feroz} {et~al.}(2019){Feroz}, {Hobson}, {Cameron}, \& {Pettitt}}]{Feroz2013}
{Feroz}, F., {Hobson}, M.~P., {Cameron}, E., \& {Pettitt}, A.~N. 2019, The Open Journal of Astrophysics, 2, 10

\bibitem[{{Fomalont} \& {Sanders}(1989)}]{1989AJ.....98..279F}
{Fomalont}, E.~B. \& {Sanders}, W.~L. 1989, \aj, 98, 279

\bibitem[{{Foreman-Mackey} {et~al.}(2013){Foreman-Mackey}, {Hogg}, {Lang}, \& {Goodman}}]{ForemanMackey2013}
{Foreman-Mackey}, D., {Hogg}, D.~W., {Lang}, D., \& {Goodman}, J. 2013, \pasp, 125, 306

\bibitem[{{Gaia Collaboration} {et~al.}(2021){Gaia Collaboration}, {Brown}, {Vallenari}, {Prusti}, {de Bruijne}, {Babusiaux}, {Biermann}, {Creevey}, {Evans}, {Eyer}, {Hutton}, {Jansen}, {Jordi}, {Klioner}, {Lammers}, {Lindegren}, {Luri}, {Mignard}, {Panem}, {Pourbaix}, {Randich}, {Sartoretti}, {Soubiran}, {Walton}, {Arenou}, {Bailer-Jones}, {Bastian}, {Cropper}, {Drimmel}, {Katz}, {Lattanzi}, {van Leeuwen}, {Bakker}, {Cacciari}, {Casta{\~n}eda}, {De Angeli}, {Ducourant}, {Fabricius}, {Fouesneau}, {Fr{\'e}mat}, {Guerra}, {Guerrier}, {Guiraud}, {Jean-Antoine Piccolo}, {Masana}, {Messineo}, {Mowlavi}, {Nicolas}, {Nienartowicz}, {Pailler}, {Panuzzo}, {Riclet}, {Roux}, {Seabroke}, {Sordo}, {Tanga}, {Th{\'e}venin}, {Gracia-Abril}, {Portell}, {Teyssier}, {Altmann}, {Andrae}, {Bellas-Velidis}, {Benson}, {Berthier}, {Blomme}, {Brugaletta}, {Burgess}, {Busso}, {Carry}, {Cellino}, {Cheek}, {Clementini}, {Damerdji}, {Davidson}, {Delchambre}, {Dell'Oro}, {Fern{\'a}ndez-Hern{\'a}ndez}, {Galluccio}, {Garc{\'\i}a-Lario},
  {Garcia-Reinaldos}, {Gonz{\'a}lez-N{\'u}{\~n}ez}, {Gosset}, {Haigron}, {Halbwachs}, {Hambly}, {Harrison}, {Hatzidimitriou}, {Heiter}, {Hern{\'a}ndez}, {Hestroffer}, {Hodgkin}, {Holl}, {Jan{\ss}en}, {Jevardat de Fombelle}, {Jordan}, {Krone-Martins}, {Lanzafame}, {L{\"o}ffler}, {Lorca}, {Manteiga}, {Marchal}, {Marrese}, {Moitinho}, {Mora}, {Muinonen}, {Osborne}, {Pancino}, {Pauwels}, {Petit}, {Recio-Blanco}, {Richards}, {Riello}, {Rimoldini}, {Robin}, {Roegiers}, {Rybizki}, {Sarro}, {Siopis}, {Smith}, {Sozzetti}, {Ulla}, {Utrilla}, {van Leeuwen}, {van Reeven}, {Abbas}, {Abreu Aramburu}, {Accart}, {Aerts}, {Aguado}, {Ajaj}, {Altavilla}, {{\'A}lvarez}, {{\'A}lvarez Cid-Fuentes}, {Alves}, {Anderson}, {Anglada Varela}, {Antoja}, {Audard}, {Baines}, {Baker}, {Balaguer-N{\'u}{\~n}ez}, {Balbinot}, {Balog}, {Barache}, {Barbato}, {Barros}, {Barstow}, {Bartolom{\'e}}, {Bassilana}, {Bauchet}, {Baudesson-Stella}, {Becciani}, {Bellazzini}, {Bernet}, {Bertone}, {Bianchi}, {Blanco-Cuaresma}, {Boch}, {Bombrun}, {Bossini},
  {Bouquillon}, {Bragaglia}, {Bramante}, {Breedt}, {Bressan}, {Brouillet}, {Bucciarelli}, {Burlacu}, {Busonero}, {Butkevich}, {Buzzi}, {Caffau}, {Cancelliere}, {C{\'a}novas}, {Cantat-Gaudin}, {Carballo}, {Carlucci}, {Carnerero}, {Carrasco}, {Casamiquela}, {Castellani}, {Castro-Ginard}, {Castro Sampol}, {Chaoul}, {Charlot}, {Chemin}, {Chiavassa}, {Cioni}, {Comoretto}, {Cooper}, {Cornez}, {Cowell}, {Crifo}, {Crosta}, {Crowley}, {Dafonte}, {Dapergolas}, {David}, {David}, {de Laverny}, {De Luise}, {De March}, {De Ridder}, {de Souza}, {de Teodoro}, {de Torres}, {del Peloso}, {del Pozo}, {Delbo}, {Delgado}, {Delgado}, {Delisle}, {Di Matteo}, {Diakite}, {Diener}, {Distefano}, {Dolding}, {Eappachen}, {Edvardsson}, {Enke}, {Esquej}, {Fabre}, {Fabrizio}, {Faigler}, {Fedorets}, {Fernique}, {Fienga}, {Figueras}, {Fouron}, {Fragkoudi}, {Fraile}, {Franke}, {Gai}, {Garabato}, {Garcia-Gutierrez}, {Garc{\'\i}a-Torres}, {Garofalo}, {Gavras}, {Gerlach}, {Geyer}, {Giacobbe}, {Gilmore}, {Girona}, {Giuffrida}, {Gomel}, {Gomez},
  {Gonzalez-Santamaria}, {Gonz{\'a}lez-Vidal}, {Granvik}, {Guti{\'e}rrez-S{\'a}nchez}, {Guy}, {Hauser}, {Haywood}, {Helmi}, {Hidalgo}, {Hilger}, {H{\l}adczuk}, {Hobbs}, {Holland}, {Huckle}, {Jasniewicz}, {Jonker}, {Juaristi Campillo}, {Julbe}, {Karbevska}, {Kervella}, {Khanna}, {Kochoska}, {Kontizas}, {Kordopatis}, {Korn}, {Kostrzewa-Rutkowska}, {Kruszy{\'n}ska}, {Lambert}, {Lanza}, {Lasne}, {Le Campion}, {Le Fustec}, {Lebreton}, {Lebzelter}, {Leccia}, {Leclerc}, {Lecoeur-Taibi}, {Liao}, {Licata}, {Lindstr{\o}m}, {Lister}, {Livanou}, {Lobel}, {Madrero Pardo}, {Managau}, {Mann}, {Marchant}, {Marconi}, {Marcos Santos}, {Marinoni}, {Marocco}, {Marshall}, {Martin Polo}, {Mart{\'\i}n-Fleitas}, {Masip}, {Massari}, {Mastrobuono-Battisti}, {Mazeh}, {McMillan}, {Messina}, {Michalik}, {Millar}, {Mints}, {Molina}, {Molinaro}, {Moln{\'a}r}, {Montegriffo}, {Mor}, {Morbidelli}, {Morel}, {Morris}, {Mulone}, {Munoz}, {Muraveva}, {Murphy}, {Musella}, {Noval}, {Ord{\'e}novic}, {Orr{\`u}}, {Osinde}, {Pagani}, {Pagano},
  {Palaversa}, {Palicio}, {Panahi}, {Pawlak}, {Pe{\~n}alosa Esteller}, {Penttil{\"a}}, {Piersimoni}, {Pineau}, {Plachy}, {Plum}, {Poggio}, {Poretti}, {Poujoulet}, {Pr{\v{s}}a}, {Pulone}, {Racero}, {Ragaini}, {Rainer}, {Raiteri}, {Rambaux}, {Ramos}, {Ramos-Lerate}, {Re Fiorentin}, {Regibo}, {Reyl{\'e}}, {Ripepi}, {Riva}, {Rixon}, {Robichon}, {Robin}, {Roelens}, {Rohrbasser}, {Romero-G{\'o}mez}, {Rowell}, {Royer}, {Rybicki}, {Sadowski}, {Sagrist{\`a} Sell{\'e}s}, {Sahlmann}, {Salgado}, {Salguero}, {Samaras}, {Sanchez Gimenez}, {Sanna}, {Santove{\~n}a}, {Sarasso}, {Schultheis}, {Sciacca}, {Segol}, {Segovia}, {S{\'e}gransan}, {Semeux}, {Shahaf}, {Siddiqui}, {Siebert}, {Siltala}, {Slezak}, {Smart}, {Solano}, {Solitro}, {Souami}, {Souchay}, {Spagna}, {Spoto}, {Steele}, {Steidelm{\"u}ller}, {Stephenson}, {S{\"u}veges}, {Szabados}, {Szegedi-Elek}, {Taris}, {Tauran}, {Taylor}, {Teixeira}, {Thuillot}, {Tonello}, {Torra}, {Torra}, {Turon}, {Unger}, {Vaillant}, {van Dillen}, {Vanel}, {Vecchiato}, {Viala}, {Vicente},
  {Voutsinas}, {Weiler}, {Wevers}, {Wyrzykowski}, {Yoldas}, {Yvard}, {Zhao}, {Zorec}, {Zucker}, {Zurbach}, \& {Zwitter}}]{gaia}
{Gaia Collaboration}, {Brown}, A.~G.~A., {Vallenari}, A., {et~al.} 2021, \aap, 649, A1

\bibitem[{{Gehrels}(1986)}]{1986ApJ...303..336G}
{Gehrels}, N. 1986, \apj, 303, 336

\bibitem[{{Gilbert} {et~al.}(2022){Gilbert}, {Barclay}, {Quintana}, {Walkowicz}, {Vega}, {Schlieder}, {Monsue}, {Cale}, {Collins}, {Gaidos}, {El Mufti}, {Reefe}, {Plavchan}, {Tanner}, {Wittenmyer}, {Wittrock}, {Jenkins}, {Latham}, {Ricker}, {Rose}, {Seager}, {Vanderspek}, \& {Winn}}]{gilbert2022}
{Gilbert}, E.~A., {Barclay}, T., {Quintana}, E.~V., {et~al.} 2022, \aj, 163, 147

\bibitem[{{G{\"u}del}(2002)}]{Gudel2002}
{G{\"u}del}, M. 2002, \araa, 40, 217

\bibitem[{{G{\"u}nther} {et~al.}(2020){G{\"u}nther}, {Zhan}, {Seager}, {Rimmer}, {Ranjan}, {Stassun}, {Oelkers}, {Daylan}, {Newton}, {Kristiansen}, {Olah}, {Gillen}, {Rappaport}, {Ricker}, {Vanderspek}, {Latham}, {Winn}, {Jenkins}, {Glidden}, {Fausnaugh}, {Levine}, {Dittmann}, {Quinn}, {Krishnamurthy}, \& {Ting}}]{2020AJ....159...60G}
{G{\"u}nther}, M.~N., {Zhan}, Z., {Seager}, S., {et~al.} 2020, \aj, 159, 60

\bibitem[{{Hallinan} {et~al.}(2015){Hallinan}, {Littlefair}, {Cotter}, {Bourke}, {Harding}, {Pineda}, {Butler}, {Golden}, {Basri}, {Doyle}, {Kao}, {Berdyugina}, {Kuznetsov}, {Rupen}, \& {Antonova}}]{hallinan2015}
{Hallinan}, G., {Littlefair}, S.~P., {Cotter}, G., {et~al.} 2015, \nat, 523, 568

\bibitem[{{Hamaker} \& {Bregman}(1996)}]{1996A&AS..117..161H}
{Hamaker}, J.~P. \& {Bregman}, J.~D. 1996, \aaps, 117, 161

\bibitem[{{Hess} \& {Zarka}(2011)}]{2011A&A...531A..29H}
{Hess}, S.~L.~G. \& {Zarka}, P. 2011, \aap, 531, A29

\bibitem[{{Hewitt} {et~al.}(1982){Hewitt}, {Melrose}, \& {Ronnmark}}]{1982AuJPh..35..447H}
{Hewitt}, R.~G., {Melrose}, D.~B., \& {Ronnmark}, K.~G. 1982, Australian Journal of Physics, 35, 447

\bibitem[{Hunter(2007)}]{Hunter:2007}
Hunter, J.~D. 2007, Computing In Science \& Engineering, 9, 90

\bibitem[{Jones {et~al.}(2001)Jones, Oliphant, Peterson, \& Others}]{scipy}
Jones, E., Oliphant, T., Peterson, P., \& Others. 2001, {SciPy}: Open source scientific tools for Python

\bibitem[{{Kao} {et~al.}(2016){Kao}, {Hallinan}, {Pineda}, {Escala}, {Burgasser}, {Bourke}, \& {Stevenson}}]{2016ApJ...818...24K}
{Kao}, M.~M., {Hallinan}, G., {Pineda}, J.~S., {et~al.} 2016, \apj, 818, 24

\bibitem[{{Kao} \& {Pineda}(2022)}]{2022ApJ...932...21K}
{Kao}, M.~M. \& {Pineda}, J.~S. 2022, \apj, 932, 21

\bibitem[{Kashyap {et~al.}(2002)Kashyap, Drake, G{\"u}del, \& Audard}]{Kashyap_2002}
Kashyap, V.~L., Drake, J.~J., G{\"u}del, M., \& Audard, M. 2002, The Astrophysical Journal, 580, 1118

\bibitem[{Kass \& Raftery(1995)}]{Kass1995}
Kass, R.~E. \& Raftery, A.~E. 1995, Journal of the American Statistical Association, 90, 773

\bibitem[{{Kavanagh} \& {Vedantham}(2023)}]{kavanagh2023}
{Kavanagh}, R.~D. \& {Vedantham}, H.~K. 2023, \mnras, 524, 6267

\bibitem[{{Kavanagh} {et~al.}(2021){Kavanagh}, {Vidotto}, {Klein}, {Jardine}, {Donati}, \& {{\'O} Fionnag{\'a}in}}]{kavanagh2021}
{Kavanagh}, R.~D., {Vidotto}, A.~A., {Klein}, B., {et~al.} 2021, \mnras, 504, 1511

\bibitem[{{Kivelson} \& {Russell}(1995)}]{kivelson95}
{Kivelson}, M.~G. \& {Russell}, C.~T. 1995, {Introduction to Space Physics}, 1st edn. (Cambridge, United Kingdom: Cambridge University Press)

\bibitem[{{Klein} {et~al.}(2022){Klein}, {Zicher}, {Kavanagh}, {Nielsen}, {Aigrain}, {Vidotto}, {Barrag{\'a}n}, {Strugarek}, {Nicholson}, {Donati}, \& {Bouvier}}]{2022MNRAS.512.5067K}
{Klein}, B., {Zicher}, N., {Kavanagh}, R.~D., {et~al.} 2022, \mnras, 512, 5067

\bibitem[{{Kochukhov} \& {Reiners}(2020)}]{2020ApJ...902...43K}
{Kochukhov}, O. \& {Reiners}, A. 2020, \apj, 902, 43

\bibitem[{{Kundu} {et~al.}(1987){Kundu}, {Jackson}, {White}, \& {Melozzi}}]{kundu1987}
{Kundu}, M.~R., {Jackson}, P.~D., {White}, S.~M., \& {Melozzi}, M. 1987, \apj, 312, 822

\bibitem[{{Kunkel}(1970)}]{1970PASP...82.1341K}
{Kunkel}, W.~E. 1970, \pasp, 82, 1341

\bibitem[{{Lazio} {et~al.}(2016){Lazio}, {Shkolnik}, {Hallinan}, \& {Planetary Habitability Study Team}}]{2016pmf..rept.....L}
{Lazio}, T. J.~W., {Shkolnik}, E., {Hallinan}, G., \& {Planetary Habitability Study Team}. 2016, {Planetary Magnetic Fields: Planetary Interiors and Habitability}, Tech. rep.

\bibitem[{{Leto} {et~al.}(2020){Leto}, {Trigilio}, {Leone}, {Pillitteri}, {Buemi}, {Fossati}, {Cavallaro}, {Oskinova}, {Ignace}, {Krti{\v{c}}ka}, {Umana}, {Catanzaro}, {Ingallinera}, {Bufano}, {Agliozzo}, {Phillips}, {Cerrigone}, {Riggi}, {Loru}, {Munari}, {Gangi}, {Giarrusso}, \& {Robrade}}]{2020MNRAS.493.4657L}
{Leto}, P., {Trigilio}, C., {Leone}, F., {et~al.} 2020, \mnras, 493, 4657

\bibitem[{{Leto} {et~al.}(2017){Leto}, {Trigilio}, {Oskinova}, {Ignace}, {Buemi}, {Umana}, {Ingallinera}, {Todt}, \& {Leone}}]{2017MNRAS.467.2820L}
{Leto}, P., {Trigilio}, C., {Oskinova}, L., {et~al.} 2017, \mnras, 467, 2820

\bibitem[{{Leto} {et~al.}(2019){Leto}, {Trigilio}, {Oskinova}, {Ignace}, {Buemi}, {Umana}, {Cavallaro}, {Ingallinera}, {Bufano}, {Phillips}, {Agliozzo}, {Cerrigone}, {Todt}, {Riggi}, \& {Leone}}]{2019MNRAS.482L...4L}
{Leto}, P., {Trigilio}, C., {Oskinova}, L.~M., {et~al.} 2019, \mnras, 482, L4

\bibitem[{{Llama} {et~al.}(2018){Llama}, {Jardine}, {Wood}, {Hallinan}, \& {Morin}}]{2018ApJ...854....7L}
{Llama}, J., {Jardine}, M.~M., {Wood}, K., {Hallinan}, G., \& {Morin}, J. 2018, \apj, 854, 7

\bibitem[{Lomb(1976)}]{lomb}
Lomb, N.~R. 1976, Astrophysics and Space Science, 39, 447

\bibitem[{{Mamajek} \& {Bell}(2014)}]{2014MNRAS.445.2169M}
{Mamajek}, E.~E. \& {Bell}, C. P.~M. 2014, \mnras, 445, 2169

\bibitem[{{Marques} {et~al.}(2017){Marques}, {Zarka}, {Echer}, {Ryabov}, {Alves}, {Denis}, \& {Coffre}}]{marques2017}
{Marques}, M.~S., {Zarka}, P., {Echer}, E., {et~al.} 2017, \aap, 604, A17

\bibitem[{{Martioli} {et~al.}(2021){Martioli}, {H{\'e}brard}, {Correia}, {Laskar}, \& {Lecavelier des Etangs}}]{aumic_c}
{Martioli}, E., {H{\'e}brard}, G., {Correia}, A.~C.~M., {Laskar}, J., \& {Lecavelier des Etangs}, A. 2021, \aap, 649, A177

\bibitem[{{McLean} \& {Labrum}(1985)}]{1985srph.book.....M}
{McLean}, D.~J. \& {Labrum}, N.~R. 1985, {Solar radiophysics : studies of emission from the sun at metre wavelengths}

\bibitem[{{Melrose} \& {Dulk}(1982)}]{1982ApJ...259..844M}
{Melrose}, D.~B. \& {Dulk}, G.~A. 1982, \apj, 259, 844

\bibitem[{{Offringa} {et~al.}(2014){Offringa}, {McKinley}, {Hurley-Walker}, {Briggs}, {Wayth}, {Kaplan}, {Bell}, {Feng}, {Neben}, {Hughes}, {Rhee}, {Murphy}, {Bhat}, {Bernardi}, {Bowman}, {Cappallo}, {Corey}, {Deshpande}, {Emrich}, {Ewall-Wice}, {Gaensler}, {Goeke}, {Greenhill}, {Hazelton}, {Hindson}, {Johnston-Hollitt}, {Jacobs}, {Kasper}, {Kratzenberg}, {Lenc}, {Lonsdale}, {Lynch}, {McWhirter}, {Mitchell}, {Morales}, {Morgan}, {Kudryavtseva}, {Oberoi}, {Ord}, {Pindor}, {Procopio}, {Prabu}, {Riding}, {Roshi}, {Shankar}, {Srivani}, {Subrahmanyan}, {Tingay}, {Waterson}, {Webster}, {Whitney}, {Williams}, \& {Williams}}]{2014MNRAS.444..606O}
{Offringa}, A.~R., {McKinley}, B., {Hurley-Walker}, N., {et~al.} 2014, \mnras, 444, 606

\bibitem[{{Osten} \& {Bastian}(2006)}]{2006ApJ...637.1016O}
{Osten}, R.~A. \& {Bastian}, T.~S. 2006, \apj, 637, 1016

\bibitem[{{Osten} \& {Bastian}(2008)}]{2008ApJ...674.1078O}
{Osten}, R.~A. \& {Bastian}, T.~S. 2008, \apj, 674, 1078

\bibitem[{Owen(2019)}]{owen2019}
Owen, J.~E. 2019, Annual Review of Earth and Planetary Sciences, 47, 67

\bibitem[{{Pallavicini} {et~al.}(1990){Pallavicini}, {Tagliaferri}, \& {Stella}}]{1990A&A...228..403P}
{Pallavicini}, R., {Tagliaferri}, G., \& {Stella}, L. 1990, \aap, 228, 403

\bibitem[{P\'erez \& Granger(2007)}]{PER-GRA:2007}
P\'erez, F. \& Granger, B.~E. 2007, Computing in Science and Engineering, 9, 21

\bibitem[{{P{\'e}rez-Torres} {et~al.}(2021){P{\'e}rez-Torres}, {G{\'o}mez}, {Ortiz}, {Leto}, {Anglada}, {G{\'o}mez}, {Rodr{\'\i}guez}, {Trigilio}, {Amado}, {Alberdi}, {Anglada-Escud{\'e}}, {Osorio}, {Umana}, {Berdi{\~n}as}, {L{\'o}pez-Gonz{\'a}lez}, {Morales}, {Rodr{\'\i}guez-L{\'o}pez}, \& {Chibueze}}]{perez-torres2021}
{P{\'e}rez-Torres}, M., {G{\'o}mez}, J.~F., {Ortiz}, J.~L., {et~al.} 2021, \aap, 645, A77

\bibitem[{{Pineda} \& {Villadsen}(2023)}]{pineda2023}
{Pineda}, J.~S. \& {Villadsen}, J. 2023, Nature Astronomy, 7, 569

\bibitem[{{Plavchan} {et~al.}(2020){Plavchan}, {Barclay}, {Gagn{\'e}}, {Gao}, {Cale}, {Matzko}, {Dragomir}, {Quinn}, {Feliz}, {Stassun}, {Crossfield}, {Berardo}, {Latham}, {Tieu}, {Anglada-Escud{\'e}}, {Ricker}, {Vanderspek}, {Seager}, {Winn}, {Jenkins}, {Rinehart}, {Krishnamurthy}, {Dynes}, {Doty}, {Adams}, {Afanasev}, {Beichman}, {Bottom}, {Bowler}, {Brinkworth}, {Brown}, {Cancino}, {Ciardi}, {Clampin}, {Clark}, {Collins}, {Davison}, {Foreman-Mackey}, {Furlan}, {Gaidos}, {Geneser}, {Giddens}, {Gilbert}, {Hall}, {Hellier}, {Henry}, {Horner}, {Howard}, {Huang}, {Huber}, {Kane}, {Kenworthy}, {Kielkopf}, {Kipping}, {Klenke}, {Kruse}, {Latouf}, {Lowrance}, {Mennesson}, {Mengel}, {Mills}, {Morton}, {Narita}, {Newton}, {Nishimoto}, {Okumura}, {Palle}, {Pepper}, {Quintana}, {Roberge}, {Roccatagliata}, {Schlieder}, {Tanner}, {Teske}, {Tinney}, {Vanderburg}, {von Braun}, {Walp}, {Wang}, {Wang}, {Weigand}, {White}, {Wittenmyer}, {Wright}, {Youngblood}, {Zhang}, \& {Zilberman}}]{aumic_b}
{Plavchan}, P., {Barclay}, T., {Gagn{\'e}}, J., {et~al.} 2020, \nat, 582, 497

\bibitem[{Rybicki \& Lightman(1979)}]{Rybicki_Lightman}
Rybicki, G.~B. \& Lightman, A.~P. 1979, Radiative processes in astrophysics (John Wiley \& Sons)

\bibitem[{{Saur} {et~al.}(2013){Saur}, {Grambusch}, {Duling}, {Neubauer}, \& {Simon}}]{2013A&A...552A.119S}
{Saur}, J., {Grambusch}, T., {Duling}, S., {Neubauer}, F.~M., \& {Simon}, S. 2013, \aap, 552, A119

\bibitem[{{Scaife} \& {Heald}(2012)}]{Scaife2012}
{Scaife}, A.~M.~M. \& {Heald}, G.~H. 2012, \mnras, 423, L30

\bibitem[{{Scargle}(1982)}]{scargle}
{Scargle}, J.~D. 1982, \apj, 263, 835

\bibitem[{{Torres} {et~al.}(2006){Torres}, {Quast}, {da Silva}, {de La Reza}, {Melo}, \& {Sterzik}}]{2006A&A...460..695T}
{Torres}, C.~A.~O., {Quast}, G.~R., {da Silva}, L., {et~al.} 2006, \aap, 460, 695

\bibitem[{{Treumann}(2006)}]{2006A&ARv..13..229T}
{Treumann}, R.~A. 2006, \aapr, 13, 229

\bibitem[{{Trigilio} {et~al.}(2023){Trigilio}, {Biswas}, {Leto}, {Umana}, {Busa}, {Cavallaro}, {Das}, {Chandra}, {Perez-Torres}, {Wade}, {Bordiu}, {Buemi}, {Bufano}, {Ingallinera}, {Loru}, \& {Riggi}}]{yzceti2}
{Trigilio}, C., {Biswas}, A., {Leto}, P., {et~al.} 2023, arXiv e-prints, arXiv:2305.00809

\bibitem[{Van Der~Walt {et~al.}(2011)Van Der~Walt, Colbert, \& Varoquaux}]{van2011numpy}
Van Der~Walt, S., Colbert, S.~C., \& Varoquaux, G. 2011, Computing in Science \& Engineering, 13, 22

\bibitem[{{van Diepen} {et~al.}(2018){van Diepen}, {Dijkema}, \& {Offringa}}]{2018ascl.soft04003V}
{van Diepen}, G., {Dijkema}, T.~J., \& {Offringa}, A. 2018, {DPPP: Default Pre-Processing Pipeline}

\bibitem[{{Vedantham} {et~al.}(2020){Vedantham}, {Callingham}, {Shimwell}, {Tasse}, {Pope}, {Bedell}, {Snellen}, {Best}, {Hardcastle}, {Haverkorn}, {Mechev}, {O'Sullivan}, {R{\"o}ttgering}, \& {White}}]{j1019}
{Vedantham}, H.~K., {Callingham}, J.~R., {Shimwell}, T.~W., {et~al.} 2020, Nature Astronomy, 4, 577

\bibitem[{{Villadsen} \& {Hallinan}(2019)}]{villadsen2019}
{Villadsen}, J. \& {Hallinan}, G. 2019, \apj, 871, 214

\bibitem[{{Wild} \& {McCready}(1950)}]{1950AuSRA...3..387W}
{Wild}, J.~P. \& {McCready}, L.~L. 1950, Australian Journal of Scientific Research A Physical Sciences, 3, 387

\bibitem[{{Wilson} {et~al.}(2011){Wilson}, {Ferris}, {Axtens}, {Brown}, {Davis}, {Hampson}, {Leach}, {Roberts}, {Saunders}, {Koribalski}, {Caswell}, {Lenc}, {Stevens}, {Voronkov}, {Wieringa}, {Brooks}, {Edwards}, {Ekers}, {Emonts}, {Hindson}, {Johnston}, {Maddison}, {Mahony}, {Malu}, {Massardi}, {Mao}, {McConnell}, {Norris}, {Schnitzeler}, {Subrahmanyan}, {Urquhart}, {Thompson}, \& {Wark}}]{Wilson2011}
{Wilson}, W.~E., {Ferris}, R.~H., {Axtens}, P., {et~al.} 2011, \mnras, 416, 832

\bibitem[{{Wittrock} {et~al.}(2023){Wittrock}, {Plavchan}, {Cale}, {Barclay}, {Ludwig}, {Schwarz}, {M{\'e}karnia}, {Triaud}, {Abe}, {Suarez}, {Guillot}, {Conti}, {Collins}, {Waite}, {Kielkopf}, {Collins}, {Dreizler}, {El Mufti}, {Feliz}, {Gaidos}, {Geneser}, {Horne}, {Kane}, {Lowrance}, {Martioli}, {Radford}, {Reefe}, {Roccatagliata}, {Shporer}, {Stassun}, {Stockdale}, {Tan}, {Tanner}, \& {Vega}}]{ttvs_aumic}
{Wittrock}, J.~M., {Plavchan}, P.~P., {Cale}, B.~L., {et~al.} 2023, \aj, 166, 232

\bibitem[{{Yu} {et~al.}(2023){Yu}, {Chen}, {Sharma}, {Bastian}, {Mondal}, {Gary}, {Luo}, \& {Battaglia}}]{sunspot}
{Yu}, S., {Chen}, B., {Sharma}, R., {et~al.} 2023, Nature Astronomy [\eprint[arXiv]{2310.01240}]

\bibitem[{{Zarka}(1998)}]{1998JGR...10320159Z}
{Zarka}, P. 1998, \jgr, 103, 20159

\bibitem[{{Zarka}(2007)}]{2007P&SS...55..598Z}
{Zarka}, P. 2007, \planss, 55, 598

\bibitem[{Zic {et~al.}(2020)Zic, Murphy, Lynch, Heald, Lenc, Kaplan, Cairns, Coward, Gendre, Johnston, MacGregor, Price, \& Wheatland}]{Zic_2020}
Zic, A., Murphy, T., Lynch, C., {et~al.} 2020, The Astrophysical Journal, 905, 23

\bibitem[{{Zic} {et~al.}(2019){Zic}, {Stewart}, {Lenc}, {Murphy}, {Lynch}, {Kaplan}, {Hotan}, {Anderson}, {Bunton}, {Chippendale}, {Mader}, \& {Phillips}}]{zic2019}
{Zic}, A., {Stewart}, A., {Lenc}, E., {et~al.} 2019, \mnras, 488, 559

\end{thebibliography}

\appendix

\section{Radio emission from an auroral ring}
\label{app:aur}

Here, we explore if the morphology of AU\,Mic's radio light curve shown in Fig.~\ref{fig:wrapped_lightcurve} can be explained by an auroral ring emitting near the star's magnetic poles. We first assume a dipolar configuration for the magnetic field lines on which electrons are accelerated to produce the emission. The star's rotation axis $\hat{z}_\star$ is inclined relative to the line of sight $\hat{x}$ by the angle $i_\star$. The magnetic axis of the dipole $\hat{z}_\mathrm{B}$ forms the angle $\beta$ with the rotation axis, known as the magnetic obliquity. This causes the magnetic axis to precess about the rotation axis as the star rotates. Here, $\hat{z}_\mathrm{B}$ points towards the northern magnetic pole. We note that the definitions for the vectors used here are the same as described in \citet{kavanagh2023} unless stated otherwise.

\begin{figure}
\centering
\includegraphics[width = \columnwidth]{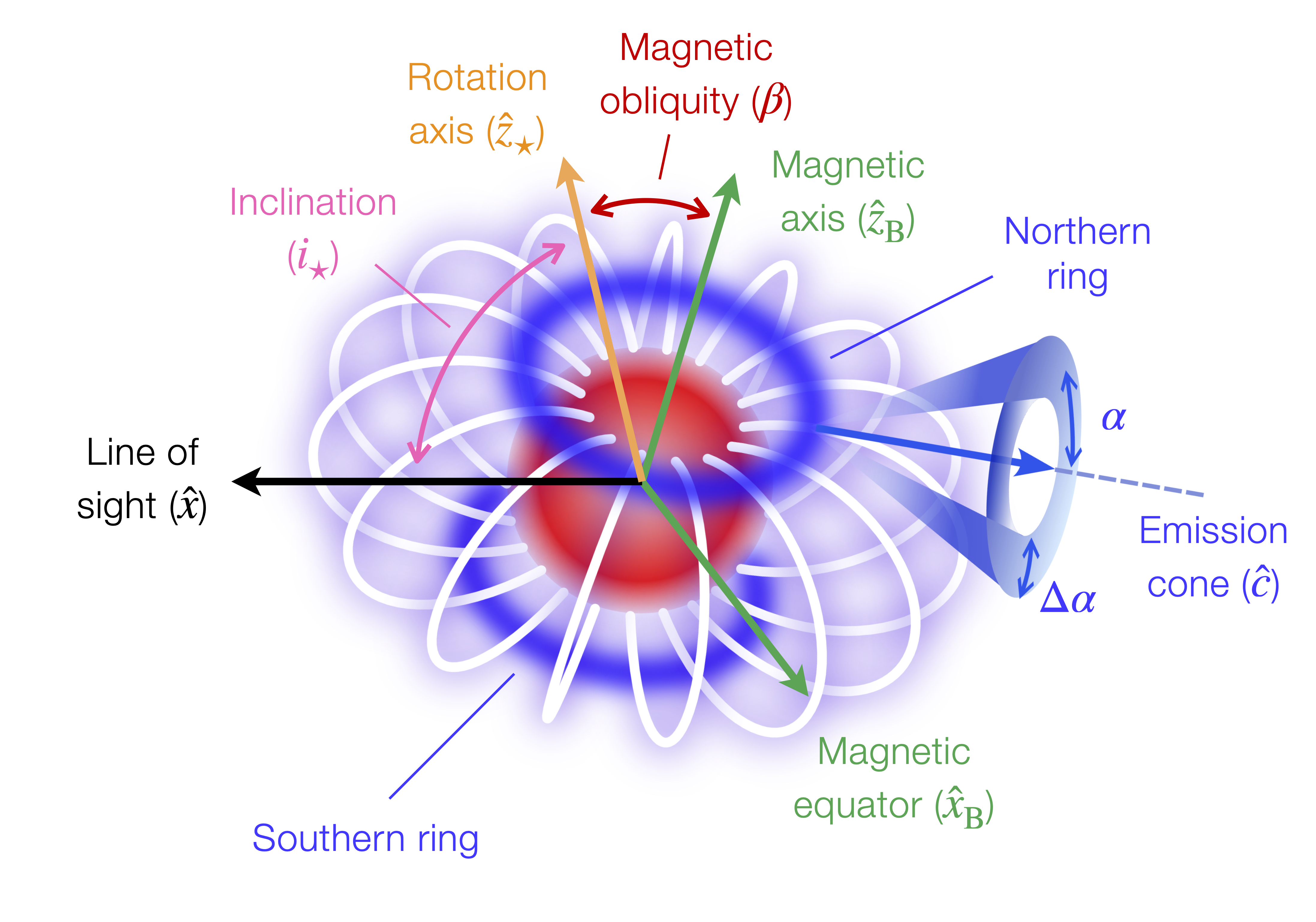}
\caption{Sketch of the geometry used to compute the theoretical light curve of an auroral ring on AU Mic. The source of the emission is assumed to be due to the electron cyclotron maser instability (ECMI) in a ring-like configuration as what is seen on Jupiter. Since ECMI is beamed, only a fraction of the auroral ring is visible at any given time. Depending on the geometry, the fraction visible changes as a function of time, modulating the relative brightness of the ring.}
\label{fig:aurora}
\end{figure}

If the observed emission is due to ECMI, it is emitted in a hollow cone \citep{dulk1985}. This cone opens outwards from the unit vector $\hat{c}$, which is parallel to the local magnetic field vector $\vec{B}$ in the northern magnetic hemisphere, and anti-parallel in the southern magnetic hemisphere. The cone has a characteristic opening angle of $\alpha$ and thickness of $\Delta\alpha$. We now assume that the observed emission comes from two rings at co-latitudes $\theta$ and $\pi-\theta$ measured from the northern magnetic pole. There is a point on both rings threaded by a single dipolar field line, which connects to the magnetic equator at the longitude $\phi_l$. Every point on each ring also has an associated emission cone. A sketch of the geometry described here is shown in Fig.~\ref{fig:aurora}.

The magnetic field vector at a point on a dipolar field line at a radius $r$ and magnetic co-latitude $\theta$ is \citep{kivelson95}
\begin{equation}
\vec{B} = B_r \hat{r} + B_\theta \hat{\theta} ,
\end{equation}
where
\begin{equation}
B_r = B_\star \Big(\frac{R_\star}{r}\Big)^3 \cos\theta ,
\end{equation}
and
\begin{equation}
B_\theta = \frac{B_\star}{2} \Big(\frac{R_\star}{r}\Big)^3 \sin\theta .
\end{equation}
Here $B_\star$ is the field strength at the magnetic poles, $R_\star$ is the stellar radius, and $\hat{r}$ and $\hat{\theta}$ are the unit vectors in the radial and meridional directions respectively. These two vectors can be expressed in terms of the magnetic axis and $\hat{x}_\mathrm{B}$, the vector that points to where the field line connects to the equator:
\begin{align}
& \hat{r} = \sin\theta\hat{x}_\mathrm{B} + \cos\theta\hat{z}_\mathrm{B} , \\
& \hat{\theta} = \cos\theta\hat{x}_\mathrm{B} - \sin\theta\hat{z}_\mathrm{B} .
\end{align}
We note that since each field line is symmetric about the magnetic axis, the field vector has no azimuthal component. The magnitude of the vector $\vec{B}$ is
\begin{equation}
B = \frac{B_\star}{2} \Big(\frac{R_\star}{r}\Big)^3 (1 + 3\cos^2\theta)^{1/2} .
\label{eq:dipole field strength}
\end{equation}
The emission cone direction in the northern magnetic hemisphere can be then shown to be
\begin{equation}
\begin{split}
\hat{c}_\mathrm{N} & = \frac{\vec{B}}{B} = \frac{3\sin\theta\cos\theta}{(1 + 3\cos^2\theta)^{1/2}}\hat{x}_\mathrm{B} + \frac{3\cos^2\theta - 1}{(1 + 3\cos^2\theta)^{1/2}} \hat{z}_\mathrm{B} \\
& = a(\theta)\hat{x}_\mathrm{B} + b(\theta)\hat{z}_\mathrm{B} .
\end{split}
\end{equation}
The direction of a cone on the southern ring on the other hand is
\begin{equation}
\hat{c}_\mathrm{S} = \frac{-\vec{B}}{B}(\pi-\theta) = a(\theta)\hat{x}_\mathrm{B} - b(\theta)\hat{z}_\mathrm{B} .
\end{equation}

Each emission cone forms the angle $\gamma$ with the line of sight. If this angle is within the range $\alpha\pm\Delta\alpha/2$, then it is visible to the observer. For a cone on the northern ring, we have
\begin{equation}
\cos\gamma = a(\theta)\hat{x}_\mathrm{B}\cdot\hat{x} + b(\theta)\hat{z}_\mathrm{B}\cdot\hat{x} .
\label{eq:beam angle}
\end{equation}
The term $\hat{x}_\mathrm{B}\cdot\hat{x}$ is
\begin{equation}
\hat{x}_\mathrm{B}\cdot\hat{x} = d(\phi_\star)\sin\phi_l + e(\phi_\star)\cos\phi_l ,
\end{equation}
where
\begin{align}
& d(\phi_\star) = -\sin i_\star \sin\phi_\star , \\
& e(\phi_\star) = \sin i_\star\cos\beta\cos\phi_\star - \cos i_\star\sin\beta ,
\end{align}
and $\phi_\star$ is the stellar rotation phase. Similarly, the term $\hat{z}_\mathrm{B}\cdot\hat{x}$ is
\begin{equation}
\hat{z}_{\mathrm{B}}\cdot\hat{x} = f(\phi_\star) = \sin i_\star \sin\beta\cos\phi_\star + \cos i_\star\cos\beta .
\end{equation}
Plugging everything in to Eq.~\ref{eq:beam angle}, we then have
\begin{equation}
\cos\gamma = g(\theta,\phi_\star)\sin\phi_l + f(\theta,\phi_\star)\cos\phi_l + i(\theta, \phi_\star),
\label{eq:beam angle N}
\end{equation}
where $g(\theta,\phi_\star) = a(\theta)d(\phi_\star)$, $h(\theta,\phi_\star) = a(\theta)e(\phi_\star)$, and $i(\theta,\phi_\star) = b(\theta)f(\phi_\star)$. For the southern ring, one can show that
\begin{equation}
\cos\gamma = g(\theta,\phi_\star)\sin\phi_l + f(\theta,\phi_\star)\cos\phi_l - i(\theta, \phi_\star).
\label{eq:beam angle S}
\end{equation}

For a given inclination $i_\star$ and magnetic obliquity $\beta$, we analytically solve Eqs.~\ref{eq:beam angle N} and \ref{eq:beam angle S} for the range of longitudes $\Delta\phi$ where $\gamma$ is in the range $\alpha \pm \Delta\alpha/2$ for each magnetic hemisphere as a function of rotation phase $\phi_\star$.
The probability of detecting bursts in this model is proportional to the fraction of the ring where the emission cones are visible. Since each point on the ring connects to the magnetic equator at a longitude $\phi_l$, the range of line longitudes $\Delta\phi$ gives the visible fraction of each ring. 

\label{lastpage}

\end{document}